\documentclass[aps,pra,twocolumn,amssymb]{revtex4-1}

\usepackage{bm}
\usepackage{graphics}
\usepackage[dvips]{graphicx}
\usepackage{float,graphicx}
\usepackage{amsmath}
\usepackage{epstopdf}
\usepackage{hyperref}
\usepackage{array,longtable}
\usepackage{multirow}
\hypersetup{
colorlinks,
citecolor=blue,
filecolor=blue,
linkcolor=blue,
urlcolor=blue, 
pdfborder = {0 0 0}, 
linktocpage
}

\newcommand{\abs}[1]{\left\vert #1\right\vert}

\def\d{\delta}
\def\wt{\widetilde}
\def\lb{\label}

\def\dag{^{\dagger}}
\def\eg{e.g.~}
\newcommand{\eq}[1]{\begin{align} #1 \end{align}}
\def\oc{Ref.~\onlinecite}
\newcommand{\bra}[1]{\langle #1 \vert} 
\newcommand{\ket}[1]{\vert #1 \rangle} 
\newcommand{\ip}[2]{\langle #1 \vert #2 \rangle}  

\newcommand{\tl}{_\tx}
\renewcommand{\th}{^\tx}

\newcommand{\beq}{\begin{equation}}
\newcommand{\eeq}{\end{equation}}
\newcommand{\beqa}{\begin{eqnarray}}
\newcommand{\eeqa}{\end{eqnarray}}

\newcommand{\tit}{\textit}
\newcommand{\tx}{\text}

\def\be{\begin{equation}}

\def\ee{\end{equation}}

\def\mmax{\textmd{max}}

\def\Rre{\textmd{Re}}
\def\Iim{\textmd{Im}}

\def\DDOS{\textmd{DOS}}

\def\be{\begin{equation}}
\def\ee{\end{equation}}
\def\bea{\begin{eqnarray}}
\def\eea{\end{eqnarray}}
\def\bit{\begin{itemize}}
\def\eit{\end{itemize}}

\begin{document}
\title{How to discretize a quantum bath for real-time evolution}
\author{In{\'e}s de Vega}
\affiliation{Department of Physics and Arnold Sommerfeld Center for Theoretical Physics, Ludwig-Maximilians-Universit{\"a}t M{\"u}nchen, Theresienstr. 37, 80333 Munich, Germany }
\author{Ulrich Schollw\"ock}
\affiliation{Department of Physics and Arnold Sommerfeld Center for Theoretical Physics, Ludwig-Maximilians-Universit{\"a}t M{\"u}nchen, Theresienstr. 37, 80333 Munich, Germany }
\author{F.~Alexander Wolf}
\affiliation{Department of Physics and Arnold Sommerfeld Center for Theoretical Physics, Ludwig-Maximilians-Universit{\"a}t M{\"u}nchen, Theresienstr. 37, 80333 Munich, Germany }

\begin{abstract}
Many numerical techniques for the description of quantum
systems that are coupled to a continuous bath 
require the discretization of the latter.
To this end, a wealth of methods has been developed in the literature, 
which we classify as (i) direct discretization, (ii) orthogonal polynomial,  
and (iii) numerical optimization strategies.
We recapitulate strategies (i) and (ii) to clarify their relation.
For quadratic Hamiltonians, we show that (ii) is the best strategy in the sense that it gives the numerically exact time evolution up to a maximum time $t_\text{max}$, for which we give a simple expression. For non-quadratic Hamiltonians, we show that no such best strategy exists. We present numerical examples relevant to open quantum systems
and strongly correlated systems, as treated by dynamical mean-field theory (DMFT).
\end{abstract}
\maketitle

\section{Introduction}

Quantum systems coupled to a continuous bath  
appear in different fields of physics, such as 
open quantum systems (OQS), 
strongly correlated many-body physics,  
and spectroscopy and scattering problems.
In the context of OQS \cite{breuerbook,rivas2011a}, 
for instance, a quantum system like an atom or a quantum dot 
is linearly coupled to a continuous bath 
like a phononic, electronic or photonic reservoir, 
which produces dissipation and decoherence in the system. 
In the context of strongly correlated many-body physics,  
the Anderson impurity model \cite{anderson1961} and its generalizations, 
which describe clusters of electronic impurities
coupled to a continuous conduction band of electrons,
are an important field of study.  
In addition, they are the basis
for dynamical mean-field theory (DMFT) \cite{metzner89,georges92,georges1996},  
which is the most widely used numerical method to describe strongly correlated systems 
in dimensions higher than one in physics \cite{kotliar06,maier05} and is popular also in quantum chemistry \cite{zgid10}.
A discrete system coupled to a continuum appears also in spectroscopy or scattering problems \cite{domcke1991}, 
leading to a \textit{resonance} or state with a complex energy that due to the imaginary energy component decays in time.

The dynamics of a system that is strongly coupled to a continuous environment
cannot be described using analytic weak-coupling approaches \cite{breuerbook,rivas2011a}, 
and requires the use of numerical techniques such as exact diagonalization (ED), the density matrix 
renormalization group (DMRG) and the numerical renormalization group (NRG). 
However, all of these numerical techniques are restricted to treating \tit{discrete} Hamiltonians,
and cannot \textit{directly} deal with a Hamiltonian that involves a continuous bath. Therefore, it is necessary to 
construct a discrete approximation to the continuous Hamiltonian.

In this paper, we analyze the problem of constructing the discrete Hamiltonian that best approximates the time evolution produced by the continuous Hamiltonian with the smallest possible number, $N_b$, of discrete degrees of freedom. As the many-body Hilbert space grows exponentially with $N_b$, this question is highly relevant, and its solution would allow to tackle systems with a complexity that is otherwise out of reach. 
We will show that this problem can only be solved for quadratic Hamiltonians. For non-quadratic Hamiltonians, we show that no
\tit{best discrete approximation} exists, and instead, heuristic arguments have to be used
to construct an approximation, as already found frequently in the literature \cite{heidrich-meisner09,peters11,guettge12,ganahl14,wolf2014,weichselbaum09,wolf2015,zwolak2008}.

Let us consider a general setup consisting of a system with Hamiltonian $H\tl{sys}$
expressed in terms of system operators $d^\dagger$ and $d$ (e.g.~in the 
quadratic case $H\tl{sys} = \epsilon_0 d^\dagger d$), which is 
linearly coupled to a continuous harmonic oscillator bath characterized by a Hamiltonian $H\tl{bath}$, 
\begin{subequations}
\begin{align}
H                   = & \, H\tl{sys}+ H\tl{bath} + H\tl{coupl}, \label{Hcont}\\
H\tl{bath}     =& \int_{a}^{b} dx\,x\,a_x\dag a_x, \label{Hcont1}\\
H\tl{coupl}   = &\int_{a}^{b} dx\,V(x) d\dag a_x + \text{h.c.}, 
\end{align}
\end{subequations}
via a ``coupling function" $V(x)$.  Here, $a_x^\dagger$ ($a_x$) create (annihilate) an occupation of a bath level 
with energy $x$.
This defines the bath spectral density $J(x)$ as \cite{breuerbook,hewson2003} 
\eq{\label{Jcont}
J(x)  = & \int_a^b dx'\,|V(x')|^2 \delta(x-x') = |V(x)|^2. 
}
This spectral density, which depends on the continuous bath variable $x$, fully characterizes the influence of the bath on the system. 
Similarly, a system linearly coupled to a discrete harmonic oscillator bath is characterized by a Hamiltonian 
\begin{subequations}
\label{Hdiscr}
\begin{align}
H\th{discr} & =H_\tx{sys}+ H\th{discr}\tl{bath}+H\th{discr}\tl{coupl} \\
H\th{discr}\tl{bath}   = & \sum_{n=1}^{N_b} x_n c_n^\dagger c_n, \\
H\th{discr}\tl{coupl} = & \sum_{n=1}^{N_b} V_n d^\dagger c_n + \tx{h.c.}.
\end{align}
\end{subequations}
The bath spectral density is a comb of delta peaks and not a continuous function as in equation \eqref{Jcont} \cite{hewson2003},
\eq{\label{Jdiscr}
J\th{discr}(x) = & \sum_{n=1}^{N_b} |V_n|^2 \delta(x-x_n).
}

For $N_b\rightarrow\infty$ one can find an $H\th{discr}$ that is 
equivalent to $H$ \cite{chinbook2011,bulla2008}. For  $N_b<\infty$, the discrete Hamiltonian 
\eqref{Hdiscr} can only serve as an approximation of the continuous Hamiltonian \eqref{Hcont}.
We classify the strategies for constructing such an approximation as follows.
\begin{itemize}
\item[(i)] 
\tit{Direct discretization},
in which bath energies $x_n$ and couplings $V_n$ are obtained
by a discretization of the integration interval $[a,b]$ in (\ref{Hcont}). 
This technique is standard in the context of NRG \cite{bulla2008} and 
frequently used in the context of DMRG \cite{heidrich-meisner09,peters11,guettge12,ganahl14,wolf2014,weichselbaum09,wolf2015,zwolak2008}.
\item[(ii)]
\tit{Orthogonal polynomials} \cite{burkey1984}, 
with which the bath energies $x_n$ are obtained 
as the zeros of a polynomial that is associated with 
a quadrature rule for the integration over the \tit{continuous bath energies} $x$.
This has been used in different contexts from DMRG
to quantum chemistry \cite{kazansky1997,karski2008,shenvi2008,prior2010,chinbook2011,prior2012,devega2015,florian2015,devega2014}.
\item[(iii)]
\tit{Numerical optimization}, which consists 
in choosing the parameters $x_n$ and $V_n$ by minimizing
a cost function \cite{caffarel1994,dorda2014,gramsch13}.
\end{itemize}

As strategy (iii) cannot be used to discretize the spectral representation of
a bath (see Appendix \ref{secOpt}), 
we restrict ourselves to strategies (i) and (ii),
which we recapitulate in Sec.~(\ref{direct}) and Sec.~(\ref{quadraturebased}), 
respectively. In Section (\ref{standard}), 
we clarify the relation of strategies (i) and (ii), which has 
hitherto been missing from the literature.
In Sec.~(\ref{timeEvol}), we show that strategy (ii)
best describes the time-evolution for quadratic Hamiltonians,
and that for non-quadratic Hamiltonians, there is no such best strategy.
Section (\ref{results}) presents numerical examples
and in Sec.~(\ref{conclusion}) we draw the main conclusions of the paper.

\section{Relation of different discretization strategies}
\label{standard}

Let us introduce the analytic continuation
of the bath spectral density \eqref{Jcont} to the complex plane,
the \tit{hybridization function} \cite{hewson2003} (see Appendix \ref{secGreen}) 
\eq{\label{Lambda}
\Lambda(z) = \int_a^b dx\, \frac{J(x)}{z-x}, \quad z \in \mathbb{C}
}
with $J(x)=|V(x)|^2$. By the Sokhotski-Plemelj theorem this implies
\eq{
J(x) = -\frac{1}{\pi} \tx{Im} \Lambda(x+i0).
}
The hybridization function does not contain more information than $J(x)$ 
since its real and imaginary parts are related by the Kramers-Kronig relation, 
$\Rre[\Lambda(x)]=\int dx \frac{\Iim[\Lambda(x)]}{x-x'}=-\frac{1}{\pi}\int dx \frac{J(x)}{x-x'}$.
Using the discrete bath spectral density $J\th{discr}(x)$ of \eqref{Jdiscr} to evaluate \eqref{Lambda}, one obtains
\begin{subequations}
\eq{
\Lambda\th{discr}(z) = & \sum^N_{n=1} \frac{|V_n|^2}{z-x_n}.  \label{LambdaDiscr}
}
\end{subequations}

\subsection{Direct discretization strategies}
\label{direct}

Let us consider the approach of \oc{shenvi2008} 
and rephrase the problem of discretizing the Hamiltonian 
as that of discretizing the integral in \eqref{Lambda}. The simplest 
approximation for an integral is obtained by using a trapezoidal integration rule
\eq{ \label{trapezoid}
\Lambda(z) = \int_a^b dx\, \frac{|V(x)|^2}{z-x} 
\simeq \sum_n \frac{|V(x_n)|^2 \Delta x_n}{z-x_n} = \Lambda\th{discr}(z),
}
where $x_n$ are linearly spaced node points with spacing $\Delta x_n$. 
Using this rule to generate an 
approximation $\Lambda\th{discr}(z)$, 
i.e.\ demanding the last equality of the preceding equation to hold,
it is possible to identify the couplings as 
\eq{\label{Vntrap}
|V_n|^2 = |V(x_n)|^2 \Delta x_n
}
and the node points $x_n$ as bath energies of \eqref{Hdiscr}.

%
%

The strategy using the trapezoidal rule can be improved as follows. Instead 
of generating a discrete weight $|V_n|^2$ simply by multiplying the function $|V(x_n)|^2$
with the width of the associated interval $\Delta x$ as in \eqref{Vntrap},  
compute the weight $|V_n|^2$ as an integral of $|V(x)|^2$ over an interval $I_n$, 
and the bath energies $x_n$ as weighted averages over this interval
\begin{subequations}
\label{Vnxndiscr}
\begin{align}
|V_n|^2 = & \int_{I_n} dx\,  |V(x)|^2, \label{Vndiscr} \\
x_n        = & \frac{1}{|V_n|^2} \int_{I_n} dx\, x\, |V(x)|^2.
\end{align}
\end{subequations}
This requires to define intervals $I_n\subset [a.b]$, $n=1,...,N_b$, 
with $I_n\cap I_m = \emptyset$ for $n\neq m$ 
and $ [a,b] \subset \bigcup_{n}  I_n$. 
For a linear discretization this generates intervals of equal width as in 
the trapezoidal rule \eqref{trapezoid}.  
But in general, the intervals $I_n$ can have arbitrary widths, 
and one can \eg define a logarithmic discretization, 
for which the interval widths decrease exponentially for $|x|\rightarrow0$.
This guarantees \tit{energy scale separation}, which is required for NRG \cite{bulla2008}. 
ED and DMRG, by contrast, allow for any discretization. 
Within DMRG, for instance, aside from 
the linear \cite{peters11,ganahl14,wolf2014} and logarithmic discretizations \cite{heidrich-meisner09,peters11}, it is possible to consider 
combinations of both discretizations \cite{weichselbaum09}, 
combinations of different logarithmic discretizations \cite{guettge12}, 
or a cosine-spaced discretization \cite{wolf2015}. Also, a parabolic discretization has been proposed \cite{zwolak2008}.

Within the direct discretization strategy, the discrete bath operators $c_n\dag$ in \eqref{Hdiscr}  
are interpreted as averages of the continuous bath operators $a_x\dag$ in \eqref{Hcont} 
over the energy interval $I_n$
\eq{\label{opdiscr}
c_{n}\dag = \frac{1}{V_n}\int_{I_n}dx \, V(x)a_x\dag.
} 
The map $a_x\dag \mapsto c_n\dag$ retains the (anti-)commutation 
relation of the continuous operators $[a_x,a_{x'}\dag]_\pm=\delta(x-x')$ 
as discretization intervals do not overlap and are normalized
\bea
[c_{n},c_{m}\dag]_\pm=\tfrac{1}{V_m^*V_n}\int_{I_n}\!\!\! dx\int_{I_m}\!\!\! dx' \, V^*(x) \, V(x')[a_x,a_{x'}\dag]_\pm = \delta_{nm}.\nonumber
\eea

In the context of direct discretization strategies, we point
out that the discrete representation \eqref{Hdiscr} is typically referred to as the 
\tit{star representation} of the discrete Hamiltonian. 
This representation is, via a standard mapping \cite{bulla2008}, 
unitarily equivalent to a one dimensional tight binding chain, 
i.e. a \tit{chain representation} (see Appendix~\ref{secLanczos}).
This mapping is valid independently of the discretization strategy and can even be formally defined to map the continuous star Hamiltonian into a chain with infinite length \cite{prior2010}. This issue will be further discussed in Sect. (\ref{chain}). 
Finally, we note that in the chain representation, the logarithmic discretization leads to 
next-neighbour couplings that decay exponentially with the distance to the impurity.

\subsubsection{New proposals}
\label{alternative}

To improve the accuracy of the discretization of previous strategies \cite{bulla2008,heidrich-meisner09,peters11,guettge12,ganahl14,wolf2014,weichselbaum09,wolf2015,zwolak2008}, 
it seems reasonable to consider a node distribution
that uses more nodes in regions where the bath spectral weight is larger. 
Based on this heuristic argument, we propose two different variants of direct discretization strategies.

In the first one, which we refer here simply as the \tit{mean method}, we compute the first bath energy as an average over the full support of $J(x)$
\eq{
x_1 & = \frac{1}{|V\tl{tot}|^2} \int_a^b dx \, x J(x),  \nonumber\\
|V\tl{tot}|^2 & = \int_a^b dx\, J(x). \label{Vtot}
}
In the next step, we compute $x_2$ as an average over
the interval $[a,x_1]$, and $x_3$ as an average over the interval $[x_1,b]$. 
The following steps are repeated in a similar way until obtaining $N_b$ energies. 
Finally, the weights $|V_n|^2$ are obtained as integrals 
 \eq{
 |V_n|^2=\int_{(x_{n-1}+x_{n})/2}^{(x_n+x_{n+1})/2}dx\,J(x),
 }
where for the first ($n=1$) and the last ($n=N_b$) integral,  
we replace the lower limit by $a$, and the upper limit by $b$, respectively.
 
Similarly, we define the \tit{equal weight method}. Here, in the first step we define a weight per bath energy 
$\frac{1}{N_b} \int_a^b dx J(x)$. 
Then, we define the first interval $I_1=[a,a_1]$ via
\eq{
\int_a^{a_1} dx\,J(x) = \frac{1}{N_b} \int_a^b dx J(x),
}
and the corresponding first bath energy and weight is computed as in (\ref{Vnxndiscr}). 
The rest of parameters $x_n$ and $V_n$ are obtained analogously.

\subsubsection{Limits of the direct discretization strategy}

The direct discretization strategies considered in this section are based on producing non-equally 
spaced discretization intervals to minimize the error of the approximation $\Lambda(z) \simeq \Lambda\th{discr}(z)$ 
for certain values of  $z=x+i0^+$, i.e.\ for certain values of the bath energy $x$. 

The logarithmic discretization, e.g., minimizes the error in the low-energy
limit $|x|\rightarrow0$. This discretization then forms a quasi-continuum in a neighborhood of $x=0$, 
and therefore the discretized version of the hybridization in such region is a numerically exact approximation to the continuous one.  However, such a good approximation for low energies comes at the price that for higher energies the discretization becomes crude, and the logarithmic approximation is therefore not appropriate 
to describe the time evolution of the system
at short and intermediate time scales. Thus, NRG, which uses 
a logarithmic discretization, allows to describe the low-energy physics of a system numerically \tit{exactly}, 
but gives a very rough approximation of high-energy excitations of the bath. 
The proposals described in Sec.~(\ref{alternative}), on the other hand, provide a good approximation in those energy regions where the spectral density is larger in magnitude, which may not necessarily coincide with low energies. 

In general, \textit{none} of the direct discretization strategies reliably describes the system at all energy scales. More precisely, a safe use of these strategies (i.e. unbiased with respect to energy) to describe time evolution at short and intermediate times scales, requires to consider a relatively high number of bath sites ($N_b=30$ up to $200$, depending on the problem \cite{heidrich-meisner09,guettge12,ganahl14,wolf2014,weichselbaum09,wolf2015,zwolak2008}).

\subsection{Orthogonal polynomial strategy}
\label{quadraturebased}

In order to construct a discrete representation of the integral (\ref{Lambda}), which is valid  
for all bath energies $x$ in $[a,b]$, it is necessary to use a discretization method 
in which each discretized energy value $x_n$ is computed with 
information of the integrand (\ref{Lambda}) over the whole integration support $[a,b]$. 
As will be described in the following, 
this can be achieved by using Gauss-Christoffel type of 
quadrature rules to represent the integral (\ref{Lambda}),
which to our knowledge has for the first time been 
proposed in Ref. \onlinecite{burkey1984}. 

\subsubsection{Gaussian quadrature}

Let us re-express the $z$-dependent
integral \eqref{Lambda} in terms of the product of a weight function $w(x)$ ($w(x)\geq0$) and a 
function $f(x,z)$ (see \oc{gautschi1981a} for an excellent review on the subject),
\eq{\label{integral}
\Lambda(z) = \int_a^b dx\, \frac{J(x)}{z-x} = \int_a^b dx\, w(x)f(x,z).
}
Now consider a polynomial interpolant $f_N(x,z)$ of $f(x,z)$ with degree $N-1$ (here
and in the following, the degree is with respect to the argument $x$, which is the integration variable),
which is unique and matches $f(x,z)$ at $N$ node points $x_n$,
\eq{
f(x,z) & = f_N(x,z) + r_N(x,z), \\
f_N(x,z) & = \sum_{n=1}^N f(x_n,z) l_n(x), \quad l_n(x_m) = \delta_{nm}, \nonumber
}
where $l_n(x)$ can be defined as the $(N-1)$-th order polynomial $l_n(x)=\prod_{m\neq n} (x-x_m)/\prod_{m\neq n} (x_n-x_m)$
and $r_N(x,z)$ is a remainder.
Clearly, if the degree of $f(x,z)$ is $N-1$, one can achieve $r_N(x,z)=0$ if choosing the $N$ node points $x_n$ \textit{intelligently},
and
\eq{\label{gausschristoffel}
\Lambda(z) & = \int_a^b dx\, w(x) f(x,z) = \sum_{n=1}^{N} W_n f(x_n,z) + R_N(z), \nonumber\\ 
W_n & = \int_a^b dx\,  w(x) l_n(x), 
}
is an \tit{exact} representation of the integral, i.e.~$R_N(z) = 0$.
We refer to $W_n$  as Christoffel weights.
It can be shown that $R_N(z) = 0$ holds even if $f(x,z)$ has a degree smaller or equal than $2N-1$,
although then $r_N(x,z)\neq 0$. The integration rule is then 
of \tit{degree of exactness} $2N-1$. The higher the degree of exactness,
the smaller is the error term $R_N(z)$ for the function $f(x,z)$, even if
the latter has degree higher than $2N-1$. 

To obtain the highest possible degree of exactness $2N-1$, 
Posse and Christoffel showed in 1877 
that the previously referred \textit{intelligent choice} of the nodes $x_n$ is to consider them as the roots of the monic polynomial $p_N(x)$ of degree $N$ 
that pertains to the family of orthogonal polynomials obeying 
\eq{\label{ortho}
\int_a^b dx\, w(x)p_n(x)p_m(x)=\delta_{nm}.
}
Such polynomials can be generated using the recurrence \cite{gautschi2005}
\eq{\label{recurrence}
p_{n+1}(x) = &  ~(x-\alpha_n) p_n(x) - \beta_n p_{n-1}(x), \\
p_{0}(x)    = & ~ 1, \qquad p_{-1}(x)    = ~0, \quad  n=0,...,N-1, \nonumber
}
where $\beta_{0} =0$ and
\begin{subequations}
\eq{
\gamma_n & = \int_a^b dx\, p_n^2(x) w(x), \\
\alpha_n & = \frac{1}{\gamma_n} \int_a^b dx\, x\, p_n^2(x) w(x), \quad  n=0,...,N-1 \\
\beta_n & = \gamma_n/\gamma_{n-1}, \quad  n=1,...,N-1.
}
\end{subequations}
It is easy to see \cite{numrec07} that the roots of $p_{N}$ can be
obtained by diagonalizing the $N\times N$ matrix $M$  \cite{golub69}
\eq{\label{jacobi}
M=\left( \begin{array}{cccc}
\alpha_0 & \sqrt{\beta_1} & 0 & \dots \\
\sqrt{\beta_1} & \alpha_1 & \sqrt{\beta_2} & \ddots \\
0 & \sqrt{\beta_2} & \alpha_2 & \ddots \\
\vdots & \ddots & \ddots & \ddots \end{array} \right).
}
In addition, denoting the $n$-th eigenvector of $M$ as $v_n$, 
the Christoffel weights in eq. (\ref{gausschristoffel}) are given by
the square of its first element
\eq{ \label{christoffelweights}
W_n = v_{n1}^2.
}
If the inner product \eqref{ortho} is not normalized, one has to multiply the right-hand side of this equation
with the norm $\int_a^b dx\, w(x)$.

\subsubsection{Discrete Hamiltonian representation}

Let us now discuss in more detail how to obtain a discrete 
Hamiltonian with $N_b$ bath sites from the $N$ roots
$x_n$, and Christoffel weights $W_n$ that appear in the Gaussian
quadrature rule for the integral \eqref{integral}. We discuss 
two cases (a) $w(x)=J(x)$ and (b) $w(x)=1$. Case (a) is, to our knowledge, 
the only one considered in the literature \cite{burkey1984,kazansky1997,karski2008,shenvi2008,prior2010,chinbook2011,prior2012,devega2015},
whereas case (b) makes the most simple choice for the weight function. 
\begin{itemize}
\item[(a)] The choice $w(x)=J(x)$
and $f_z(x)=\frac{1}{z-x}$ leads to polynomials that are orthogonal with respect to $J(x)$,
which we therefore call bath-spectral-density-orthogonal (BSDO).
Combining \eqref{integral} and \eqref{gausschristoffel} it is found
\eq{\label{eqBSDO}
\Lambda(z) \approx \sum_{n=1}^{N_b} \frac{W_n}{z-x_n} = \Lambda\th{discr}(z),
}
which allows to identify the Christoffel weights
computed via \eqref{christoffelweights}
with the weights $|V_n|^2$ of the discrete bath 
degrees of freedom
\eq{\label{christoffelW}
|V_n|^2 = W_n.
}
\item[(b)] The choice $w(x)=1$ and $f_z(x)=\frac{J(x)}{z-x}$.
This is the case of Legendre polynomials
and one obtains
\eq{\label{eqLeg}
\Lambda(z) \approx \sum_{n=1}^{N_b} \frac{W_n J(x_n)}{z-x_n} = \Lambda\th{discr}(z),
}
and the Christoffel weights $W_n$ relate to the 
weights of the discrete bath via $|V_n|^2 = W_n J(x_n)$.
\end{itemize}
  
The next question is, which of these cases leads to a better approximation?
Equations \eqref{eqBSDO} and \eqref{eqLeg} derived from \eqref{gausschristoffel} do not hold exactly:
in both cases (a) and (b) $f_z(x)$ contains a pole $\frac{1}{z-x}$
and hence it can not be \tit{exactly} represented by a polynomial of degree $2N_b-1$.
Indeed, a pole is highly difficult to approximate with polynomials and it
is quite irrelevant, whether one has an additional factor $J(x)$ that 
multiplies this pole as in case (b), if this factor $J(x)$ does not exhibit
a severe non-regular behavior. This argument is confirmed by the numerical
examples discussed in Section \ref{results}.

%
\subsubsection{Relationship to chain mappings}
\label{chain}
In this section, we show that the orthogonal polynomial method with the weight function chosen as $w(x)=J(x)$ (case (a) above), is equivalent to the \tit{chain mapping} proposed in Refs.~\onlinecite{prior2010,chinbook2011,woods2014}, and recently modified in \oc{devega2015} to tackle temperature environments in an alternative way. It is also equivalent to the chain mapping derived in the Appendix of \oc{karski2008}. 
The chain representation of the discrete \textit{star} Hamiltonian obtained by considering $w(x)=J(x)$, can be written as 
\eq{\label{map}
H\th{discr}\tl{chain}
& = H_\tx{sys} + V\tl{tot}  (d^\dagger e_0+e_0^\dagger d) \\
& \quad + \sum^{N_b-1}_{n=0} \alpha_{n} e_n^\dagger e_n + \sum^{N_b-2}_{n=0} \sqrt{\beta_{n+1}}(e^\dagger_{n+1}e_n+e_n^\dagger e_{n+1}),\nonumber
}
where $|V\tl{tot}|^2=\int_a^b dx\,J(x)$ was defined in \eqref{Vtot} and $\alpha_{n}$ and $\beta_{n}$ were defined in the recurrence relation (\ref{recurrence}).
In the limit $N_b\rightarrow \infty$, $H\th{discr}\tl{chain}$ becomes unitarily equivalent to the continuous $H$ in \eqref{Hcont},
and thus provides an \textit{exact} representation of $H$.

For finite $N_b$, the unitary transformation that takes (\ref{map}) back to its star representation (\ref{Hdiscr}), 
is equivalent to a diagonalization of the matrix (\ref{jacobi}) formed by the recurrence coefficients. As described above, such a transformation leads to the same weights and nodes as the ones obtained with the Gauss-Christoffel (BSDO quadrature). In other words, computing the system dynamics with a chain Hamiltonian (\ref{map})  is equivalent to computing the system dynamics with a star Hamiltonian (\ref{Hdiscr}) where nodes $x_n$ and weights $V_n$ are computed with the BSDO quadrature. 
Regarding the important application of DMRG calculations: in contrast to what had been commonly believed, it was only recently shown that the star representation can be much less entangled than the chain representation \cite{wolf2014b}.

Within the \tit{direct discretization strategy},
the creation operators $c_n\dag$ of the discrete Hamiltonian in the star geometry (\ref{Hdiscr}) 
were obtained as an average over the continuous bath degrees of freedom $a_x\dag$
in a small interval $I_n$, as defined in \eqref{opdiscr}. Within the \textit{orthogonal polynomial} strategy described in the current section, 
the discrete operators in the chain Hamiltonian (\ref{map}) are related to the continuous operators via
\eq{ \label{enoperators}
e_n^\dagger = \int_a^b dx\, U_n(x) a_x^\dagger,
}
where $U_{n}(x)= \sqrt{J(x)} p_{n}(x)$. 
Therefore, they correspond to a weighted average over the total support of the spectral function $J(x)$. 
Note that due to orthogonality and normalization of $p_n(x)$, the transformation is unitary 
$\int_a^b dx\, U^*_{n}(x)U_{m}(x) = \int_a^b dx\, w(x) p_n(x) p_m(x) = \delta_{nm}$ and thereby retains
the (anti-)commutation relation of $a_x\dag$. 

\subsubsection{Relationship to the Lanczos algorithm}
\label{lanzosHB}
The measure $\omega(x)=J(x)$ is commonly known as Stiltjes measure, and the three-term 
recursion \eqref{recurrence} of the associated BSDO polynomials 
is equivalent to the Lanczos algorithm for the continuous bath Hamiltonian $H\tl{bath}$ in \eqref{Hcont} \cite{gautschi2005} 
(see appendix \ref{secLanczos}). 
The environment discretization then is a consequence of truncating the
infinite recurrence relation (and therefore the matrix (\ref{jacobi})) at a finite $N=N_b$.
The implementation of the algorithm on a computer is though impossible,
as there is no direct matrix representation for the continuous $H\tl{bath}$. 

By contrast, the Lanczos algorithm is a standard procedure to tridiagonalize 
a \textit{given discrete} bath Hamiltonian $H\th{discr}\tl{bath}$ as in \eqref{Hdiscr}, 
to obtain its unitarily equivalent chain representation. In order to so, one has to 
come up with a discrete Hamiltonian in the first place, which then has to be constructed
using a \textit{direct discretization} strategy. 

\section{Time evolution}
\label{timeEvol}

\begin{table*}[ht]
\caption{Lanczos algorithm and orthogonal-polynomial strategy for real-time evolution.}
\centering
\setlength{\tabcolsep}{15pt}
\renewcommand{\arraystretch}{1.5} 
\begin{tabular}{p{0.20\linewidth}p{0.30\linewidth}p{0.30\linewidth}}
\hline
Lanczos  algorithm & Quadratic $H\tl{sys}$& Non-quadratic $H\tl{sys}$\\
\hline
For continuous $H\tl{bath}$ (eq. (\ref{Hcont1})) & ${\cal H}\th{discr}$ (eq. (\ref{fullH1})) is obtained formally (App. \ref{secLanczosGen}), and numerically (Sec. \ref{lanzosHB})  & Same as for quadratic $H\tl{sys}$ \\
For continuous $H$ (eq. (\ref{Hcont})) & ${\mathcal H}_N$ (eq. (\ref{fullH})) is obtained formally (App. \ref{secLanczosGen} and Sec. \ref{lanczosquadratic} for first steps of algorithm).& Not possible\\
Is Lanczos for $H$ equal to Lanczos for $H\tl{bath}$? & Sec. \ref{equivalence}: Yes, ${\mathcal H}_N={\cal H}\th{discr}$ for orthogonal polynomial strategy
& Sec. \ref{nonequiv}: No, in general ${\mathcal H}_N\neq{\cal H}\th{discr}$\\
\hline
\end{tabular}
\label{tabla}
\end{table*}

Let us now study the time evolution of the hybridization function, which describes
the time evolution of the bath, and the time evolution of the Green's function of the system, 
from which we can construct the time evolution of all system observables. 
The Green's function is given by 
\eq{ \label{greenEvol}
G(t) 
& = -i \bra{\psi_0} e^{-i(H-E_0)t} \ket{\psi_0}, \quad \ket{\psi_0} = d\dag \ket{E_0} \nonumber\\
& = \int_{-\infty}^{\infty} dx\,A(x) e^{-ixt},
}
where the initial state is the excitation of the system $H\tl{sys}$ through occupation with a particle,
and the spectral density of the system is
\eq{
A(x) = \sum_n |\ip{\psi_0}{E_n}|^2 \delta(x-(E_n-E_0)),
}
where the sum is over all eigenstates $\ket{E_n}$ and 
eigenenergies $E_n$ of the full Hamiltonian \eqref{Hcont}.
For a quadratic (single-particle) Hamiltonian, without loss of generality,
one can consider $E_0=0$ and $\ket{E_0}=\ket{\tx{vac}}$ and therefore
only has to study the time-evolution of a \tit{single} particle that is initially
in the system and starts interacting with the bath at non-zero times.

Analogously to \eqref{greenEvol}, we define the time evolution of the hybridization function as 
\eq{
\Lambda(t) 
& = \int_{-\infty}^{\infty} dx\,J(x) e^{-ixt}. \label{LambdaEvol}
} 
For a discrete Hamiltonian $H\th{discr}$, one obtains 
\eq{
G^\tx{discr}(t)
& = \int_{-\infty}^{\infty} dx\, A^\tx{discr}(x) e^{-i x t}, \\
\Lambda^\tx{discr}(t)
& = \sum_{n=1}^{N_b}|V_n|^2 e^{-ix_n t}.  \label{LambdaDiscrEvol}
}

In the following, it is shown that the \textit{orthogonal polynomial} strategy 
yields the best description of the short- and intermediate-time evolution
of the continuous Hamiltonian, if the latter is quadratic. It will then become
clear why none of the discretization strategies can be considered the \textit{best}
or the \textit{optimal} one if the Hamiltonian is non-quadratic (has higher order interactions). In particular:
\begin{itemize}
\item Sec.~\ref{secEvolBath} shows that the best approximation of \eqref{LambdaEvol} is obtained using the orthogonal polynomial strategy as described in Sec. \ref{quadraturebased}.
\item Sec.~\ref{lanczosquadratic} shows that the Lanzos algorithm for the full $H$ generates a matrix ${\mathcal H}_N$, which gives the nodes and the weights that approximates the Green's function \eqref{greenEvol} with a polynomial quadrature rule.
\item Sec.~\ref{equivalence} shows that if $H\tl{sys}$ is quadratic, ${\mathcal H}_N={\cal H}\th{discr}$, where ${\cal H}\th{discr}$ is obtained by Lanczos tridiagonalization of $H\tl{bath}$. Also, as it was shown in Sec. \ref{lanzosHB}, a Lanczos tridiagonalization of $H\tl{bath}$ is equivalent to a bath discretization   using the orthogonal polynomial strategy of Sec.\ref{quadraturebased}. Hence, the orthogonal polynomial strategy leads
to a quadrature rule also for the Green's function \eqref{greenEvol}.
\item Sec.~\ref{nonequiv} shows that if $H\tl{sys}$ is non-quadratic, then ${\mathcal H}_N\neq{\cal H}\th{discr}$, and nothing can be concluded about the optimality of any particular discretization method. 
\end{itemize}
An overview of these steps is provided in Table (\ref{tabla}).

\subsection{Time evolution of the bath}
\label{secEvolBath}

In Sec.~\ref{quadraturebased}, we learned that polynomial quadrature rules provide us
with the highest \textit{degree of exactness} for computing the integral \eqref{gausschristoffel}. 
In the following, we will see that this also helps us to understand in which cases \eqref{LambdaDiscrEvol} 
provides a good approximation of the Fourier type integral such as \eqref{LambdaEvol},
and how to choose the parameters of the bath in order to obtain the best approximation.
To this end, let us define the error term $R_{N_b}(t)$ and write
\eq{ \label{Lambdaevol}
\Lambda(t) = \int_{-\infty}^{\infty} dx\, J(x) e^{-ixt} = \sum_{n=1}^{N_b} |V_n|^2 e^{-ix_n t} + R_{N_b}(t).
}
We see that if we set $w(x)=J(x)$ to construct orthogonal polynomials via \eqref{recurrence}
and choose $x_n$ to be the roots of the degree $N_b$ polynomial and
$|V_n|^2 = W_n$ to be the Christoffel weights \eqref{christoffelweights}, then \eqref{Lambdaevol} 
has the form of a Gaussian quadrature rule as in \eqref{gausschristoffel} with $f(x,z=t) = e^{-ixt}$.

That is, only if we choose $|V_n|^2$ and $x_n$ according to the orthogonal polynomial
strategy with $w(x)=J(x)$, our discrete Hamiltonian corresponds to evaluating
the Fourier transform \eqref{Lambdaevol} to \textit{degree of exactness} $2N_b-1$. 
Otherwise, the degree of exactness will be lower. What does this mean in practice?

For a fixed time $t$, let us expand the part $e^{-ixt}$ of the integrand $J(x) e^{-ixt}= w(x) e^{-ixt}$ 
in \eqref{Lambdaevol} that cannot be absorbed in a weight
function in orthogonal polynomials $q_n(x)$, which are orthogonal with respect to $v(x)$ ($v(x)\geq0$ is an arbitrary weight function), according to 
\eq{ \label{expand}
e^{-ixt} 
& = \sum_{n=0}^{N} c_n q_n(x) + \sum_{n=N+1}^\infty c_n q_n(x), \nonumber \\
c_n & = \int_a^b dx\, v(x) e^{-ixt} q_n(x).
}
Let us furthermore assume the family of polynomials $q_n(x)$ to be chosen optimally for the fixed time $t$.
The optimal choice generates the most quickly converging sequence $c_n\rightarrow 0$ and by that minimizes the remainder 
$r_{N} = \sum_{n=N+1}^\infty c_n q_n(x)$ at each order of $N$. 
Of course, we don't know which polynomials these are, but this is not relevant.
The only property we need is that the coefficients become
zero for values high values of $n$: $c_n \simeq 0 \tx{ for }  n>N'(t)$, 
where $N'(t) = \tfrac{1}{2}(b-a) t$ (this is shown in Appendix \ref{secChebychev}).

The important observation to make is that choosing $x_n$ and $|V_n|^2=W_n$ 
according to the orthogonal polynomial strategy of Sec. \ref{quadraturebased}, corresponds to integrating 
the first term with $N=2N_b-1$ in \eqref{expand} exactly. Any other choice,
will lead to an exact integration of the term only at a \textit{lower} order, or will \textit{not} integrate
it exactly at \textit{any} order. Combining this observation with the fact that 
$c_n \simeq 0 \tx{ for } n > \tfrac{1}{2}(b-a) t$, we conclude that the
orthogonal polynomial strategy reproduces basically the \textit{exact}
time evolution of the hybridization function for $t<t\tl{max}$, with
\eq{\label{tmax}
t\tl{max} = 2\frac{2N_b-1}{b-a}.
}
This result is confirmed in the numerical experiments in Sec.~\ref{results}.
We have therefore shown that the best approximation of \eqref{LambdaEvol} is given by a orthogonal polynomial strategy as described in Sec. \ref{quadraturebased}.

\subsection{Time evolution of the system}

The Green's function of the system as defined in \eqref{greenEvol} can be rewritten as follows
\eq{ 
G(t) 
& = \int_{-\infty}^{\infty} dx\, A(x) e^{-ixt} = \sum_{n=1}^{\infty} |\ip{\psi_0}{E_{n-1}}|^2 e^{-iE_{n-1} t}      \nonumber \\
&= \sum_{n=1}^{N} |\ip{\psi_0}{X_n}|^2 e^{-i X_n t} + R_N(t) \label{Greenevol},
}
where $\ket{E_n}$ are eigenstates and $E_n$ eigenenergies of the exact, 
continuous Hamiltonian \eqref{Hcont}, and $R_N(t)$ is a remainder. 
The problem is therefore again to choose the states $\ket{X_n}$ and the nodes $X_n$, 
such as to make \eqref{Greenevol} a quadrature rule, which we just showed 
(Sec.~\ref{secEvolBath}) to yield the best approximation of Fourier type integrals.

\subsubsection{Lanczos for quadratic Hamiltonian}
\label{lanczosquadratic}

For quadratic Hamiltonians we will show in the following, that 
the orthogonal polynomial strategy \eqref{recurrence} 
generates a quadrature rule for \eqref{Greenevol}, 
and $X_n$ and $\ket{X_n}$ become respectively the 
eigenenergies and eigenstates of the discrete Hamiltonian $H\th{discr}$. If either one does not
use the orthogonal polynomial strategy, or the Hamitonian is not quadratic, one \textit{never}
generates a quadrature rule in \eqref{Greenevol}.

To this end, let us compute the first steps of the standard 
Lanczos tridiagonalization algorithm recapitulated in Appendix~\ref{secLanczos}. Here,
we do it for the full continuous quadratic Hamiltonian \eqref{Hcont}, and not for
the bath and coupling part of the discrete Hamiltonian \eqref{Hdiscr}, as usually done in the
context of \tit{chain mappings}.

Assume $H\tl{sys}=\varepsilon_0 d\dag d$ quadratic.  
Let us take as initial Lanczos vector the state $\ket{f_0} = \ket{d} = d\dag\ket{\tx{vac}} = \ket{\psi_0}$. 
 Denoting the single-particle states of the bath as $\ket{a_x}=a_x\dag\ket{\tx{vac}}$, we have following \eqref{lanczosGen}
\eq{
\wt \alpha_0 & = \bra{f_0} H \ket{f_0} = \varepsilon_0, \nonumber\\ 
\ket{r_0}  & = H \ket{f_0} - \wt\alpha_0 \ket{f_0} = \int_a^b dx\,V(x) \ket{a_x}, \nonumber \\
\ip{r_0}{r_0} & = \int_a^b dx\, |V(x)|^2 =  |V\tl{tot}|^2 = \wt \beta_1^2,   \nonumber \\
\ket{f_1}   & = \frac{1}{V\tl{tot}} \int_a^b dx\,V(x) \ket{a_x}. \quad  \label{LanczosHcont}
}
Continuing the algorithm up to order $N$ produces a 
truncated representation of $H$, which is a $N\times N$ matrix, 
\eq{ \label{fullH}
{\cal H}_N 
=\left( \begin{array}{ccccc}
\varepsilon_0 & V\tl{tot} & 0 & \dots \\
V\tl{tot} & \wt\alpha_1 & \sqrt{\wt\beta_2} & 0 & \dots \\
0 & \sqrt{\wt\beta_2} & \wt\alpha_2 & \sqrt{\wt\beta_3} & \ddots \\
0 & 0 & \sqrt{\wt\beta_3} & \wt\alpha_3 & \ddots \\
\vdots & \ddots & \ddots & \ddots \end{array} \right).
}
As discussed in Appendix~\ref{secLanczos}, there is a set
of orthogonal polynomials $q_n(x)$ 
that are orthogonal with respect to $w(x)=A(x)$ ($A(x)$ is the spectral density of the full Hamiltonian $H$)  
associated with the preceding Lanczos algorithm.
Therefore, diagonalization of \eqref{fullH} yields
roots $X_n$ and Christoffel weights 
$W_n = |\ip{f_0}{X_n}|^2 = |\ip{\psi_0}{X_n}|^2$. 
Hence, the Lanczos algorithm evaluated
for the continuous quadratic Hamiltonian $H$ with initial state  $\ket{f_0} = \ket{\psi_0}$ generates
the nodes and weights that make the approximation \eqref{Greenevol} a quadrature rule. Note that $X_n\neq E_n$, since $E_{n-1}$ 
are true eigenvalues of $H$, and $X_n$ are the eigenvalues of the truncated tri-diagonal representation ${\mathcal H}_N$ of $H$.

But how does this relate to the parametrization for
a discrete Hamiltonian $H\th{discr}$ that we obtain from
the orthogonal polynomial strategy \eqref{recurrence} 
for the weight function $w(x) = J(x)$? 

\subsubsection{Equivalence with orthogonal polynomial strategy}
\label{equivalence}
The discrete quadratic Hamiltonian $H\th{discr}$, 
which has dimension $(N_b+1)\times(N_b+1)$, generates the following 
approximation to the time evolution of the Green's function of the continuous system
\eq{
G(t) 
& = \int_{-\infty}^{\infty} dx\, A(x) e^{-ixt} \nonumber  \\
& =  \sum_{n=1}^{N_b+1} |\ip{\psi_0}{E_{n-1}\th{discr}}|^2 e^{-iE_{n-1}\th{discr} t} + R\th{discr}_{N_b}(t),      \label{GevolDiscr}
}
where $\ket{E_n\th{discr}}$ are eigenstates and $E_n\th{discr}$ eivenvalues of $H\th{discr}$. Also, 
$H\th{discr}$ can be represented in the chain geometry \eqref{map} as
\eq{ \label{fullH1}
{\cal H}\th{discr} 
=\left( \begin{array}{ccccc}
\varepsilon_0 & V\tl{tot} & 0 & \dots \\
V\tl{tot} & \alpha_0 & \sqrt{\beta_1} & 0 & \dots \\
0 & \sqrt{\beta_1} & \alpha_1 & \sqrt{\beta_2} & \ddots \\
0 & 0 & \sqrt{\beta_2} & \alpha_2 & \ddots \\
\vdots & \ddots & \ddots & \ddots \end{array} \right).
}
As $\ket{\psi_0}=\ket{d}$, this representation of $H\th{discr}$ directly yields the weights
and energies in \eqref{GevolDiscr}. 

In the following, we will show that the matrix \eqref{fullH1}
equals the matrix \eqref{fullH} that generates the quadrature rule, only if we compute 
the parameters of the discrete Hamiltonian using the orthogonal polynomial strategy \eqref{recurrence} 
with $w(x)=J(x)$. Only then, also \eqref{GevolDiscr} is a quadrature rule.

To this end, let us further evaluate the Lanczos algorithm for the 
continuous $H$. Using the results of \eqref{LanczosHcont}, we can represent
the terms in \eqref{Hcont} as 
\eq{
H\tl{sys} & = \varepsilon_0 \ket{f_0}\bra{f_0}, \nonumber\\
H\tl{coupl} & = V\tl{tot} (\ket{f_0}\bra{f_1} + \text{h.c.}), \nonumber\\
H\tl{bath}  & = \int_a^b dx\, x \ket{a_x}\bra{a_x}. \nonumber
}
As the Lanczos basis is orthogonal, we see that in subsequent Lanczos
steps, only $H\tl{bath}$ can contribute: $H\tl{sys}$ and $H\tl{coupl}$ 
only have contributions in the subspace spanned by $\ket{f_0}$ and $\ket{f_1}$.
We therefore have to evaluate a single next Lanczos step using the full $H$,
and from then on can iterate using only $H\tl{bath}$.
Now note that the Lanczos vector $\ket{f_1}$ in \eqref{LanczosHcont}, 
which is the starting vector for subsequent Lanczos steps, 
equals the state $\ket{e_0}$ in \eqref{eqFirstChainStateCont}, 
which is the initial state for a tridiagonalization of the bath.
We already know the latter to be equivalent to the orthogonal polynomial strategy. 
The Lanczos recursion for the full $H$ therefore generates the 
coefficients of the orthogonal polynomial strategy. 
Let us check this for the next step,
\eq{
\wt \alpha_1 & = \bra{f_1} H \ket{f_1} = \bra{f_1} H\tl{bath} \ket{f_1}. \nonumber\\ 
\ket{\wt r_1}  & = H \ket{f_1} - \wt\alpha_1 \ket{f_1} - V\tl{tot} \ket{f_0}  \nonumber \\
& = H\tl{bath} \ket{f_1} - \wt\alpha_1 \ket{f_1}. \nonumber
}
Evidently, $\wt \alpha_1 = \alpha_0$ and $\ket{\wt r_1} = \ket{r_0}$ as $\ket{f_1}=\ket{e_0}$ 
such that this equals the parameters of \eqref{lanczos} and \eqref{recurrence}. 
Hence the matrices \eqref{fullH1} and \eqref{fullH} are equivalent, and the time 
evolution computed with the discrete Hamiltonian is a quadrature rule.

For any other choice of $H\th{discr}$, which is not parametrized using 
\eqref{recurrence}, we would \textit{not} obtain an equivalent representation
to \eqref{fullH}, and therefore, \eqref{GevolDiscr} would not be a quadrature rule.

The estimate \eqref{tmax} for the maximal time $t\tl{max}$ yields, as the quadrature rule 
now uses a polynomial of degree $N_b+1$,
\eq{\label{tmaxSys}
t\tl{max} = 2\frac{2N_b+1}{b-a}.
}

\subsection{Impossibility of optimal choice for non-quadratic Hamiltonians}
\label{nonequiv}
If $H\tl{sys}$ is \textit{not} quadratic, but has higher order interaction terms, we cannot 
obtain a representation of $H\th{discr}$ in terms of single-particle states, and hence as a $(N_b+1)\times(N_b+1)$
matrix. Rather, any representation of $H\th{discr}$ then has an exponential dimension, e.g.~$2^{N_b+1}\times 2^{N_b+1}$ 
for spinless fermions, and dimension $D^{N_b+1}\times D^{N_b+1}$ for bosons with a local basis truncated at a dimension $D$. The summation over the discrete time evolution of \eqref{GevolDiscr}
then involves an exponential number of terms. 
By dimensionality, this summation can never correspond to a quadrature rule with $N_b$ parameters, which gives rise to $N_b$ roots.
The time evolution of the bath hybridization function, which always is a single-particle evolution,
is not affected by this argument and is still best described using the parameters
provided by the orthogonal polynomial strategy.

In summary, for non-quadratic Hamiltonians, even if we have a good approximation of the bath hybridization function up to $t\tl{max}$, the dynamics of the system, given by the Green function \eqref{GevolDiscr} will no longer be exact up to this time.

\section{Numerical examples}
\label{results}

\subsection{Spin-boson model}
Let us consider the Hamiltonian of an OQS with $H_\tx{sys}$ coupled to a continuous bosonic reservoir
\begin{eqnarray}
H&=&H_\tx{sys}+\int_{0}^{k_\mmax}dk\, \tilde{g}(k)\,(b(k)\sigma^{+}+\sigma^{-}b(k)^{\dagger})\nonumber\\
&+&\int_{0}^{k_\mmax}dk \, \omega(k)b(k)^{\dagger}b(k),
\label{Hamil3}
\end{eqnarray}
where $\tilde{g}(k)$ are the coupling strengths, and $b(k)$ ($b(k)^\dagger$) are harmonic oscillator operators with commutation relations $[b(k),b(k')^{\dagger}]=\delta(k-k')$. Here, the index $k$ labels the modes, which have a maximum momentum  $k_\mmax$. In the frequency representation, and provided that the environment is initially in a Gaussian state, this Hamiltonian can be rewritten as 
\begin{eqnarray}
H&=&H_\tx{sys}+\int_{0}^{\omega_\mmax}d\omega g(\omega)\,(b(\omega)\sigma^{+}+b(\omega)^{\dagger}\sigma^{-})\nonumber\\
&+&\int_{0}^{\omega_\mmax}d\omega\,  b(\omega)^{\dagger}b(\omega),
\label{Hamil3}
\eea
where $\omega_\mmax$ is determined by $k_\mmax$, and we have defined $g(\omega)=\sqrt{J(\omega)}$, where  $J(\omega)=\tilde{g}^2(\omega)\rho_\DDOS(\omega)$ is the spectral density of the environment, and $\rho_\DDOS(\omega)$ is the environment density of states.   
Hence, the Hamiltonian (\ref{Hamil3}) acquires the form (\ref{Hcont}), obviously once interpreting the continuous variable $x$ as $\omega$, and $d=\sigma^-$.
We also note that the above Hamiltonian corresponds to a simplified version of the spin-boson model, as it assumes a rotating wave approximation to discard \textit{fast rotating} terms of the form $b^\dagger(k)\sigma^+$, and $b(k)\sigma^-$. Such an approximation, which is particularly valid in quantum optics, leads to a Hamiltonian that conserves the number of particles. This simplifies considerably the numerical treatment, particularly at zero temperature.

In order to characterize the environment, let us consider a spectral density of the Caldeira and Leggett type \cite{caldeira1983,weissbook}, 
\bea
J(\omega)=\alpha\omega^s \omega_c^{1-s}e^{-\omega/\omega_c}, 
\eea
which constitute a very general description that allows to describe many different types of reservoirs, depending on the choice of the parameter $s$. The exponential factor in this model provides a smooth regularization for the spectral density, being modulated by the frequency $\omega_c$. Environments with $0<s<1$ are considered as \textit{sub-ohmic}, while those corresponding to $s=1$ and $s>1$ are known as \textit{ohmic} and \textit{super-ohmic} respectively. The constant $\alpha$ describes the coupling strength of the system and the environment.
In the following, we will focus on a sub-ohmic spectral density with $s=1/2$. Sub-ohmic spectral densities describe the frequency dependence of photonic bands in photonic band gap materials \cite{florescu2001,devega2005,prior2012}, as well as the dominant noise sources in solid state devices at low temperatures such as superconducting qubits \cite{shnirman2002}, nanomechanical oscillators \cite{seoanez2007}, and quantum dots \cite{tong2006}. 

\begin{figure}[t]
\centerline{\includegraphics[width=0.45\textwidth]{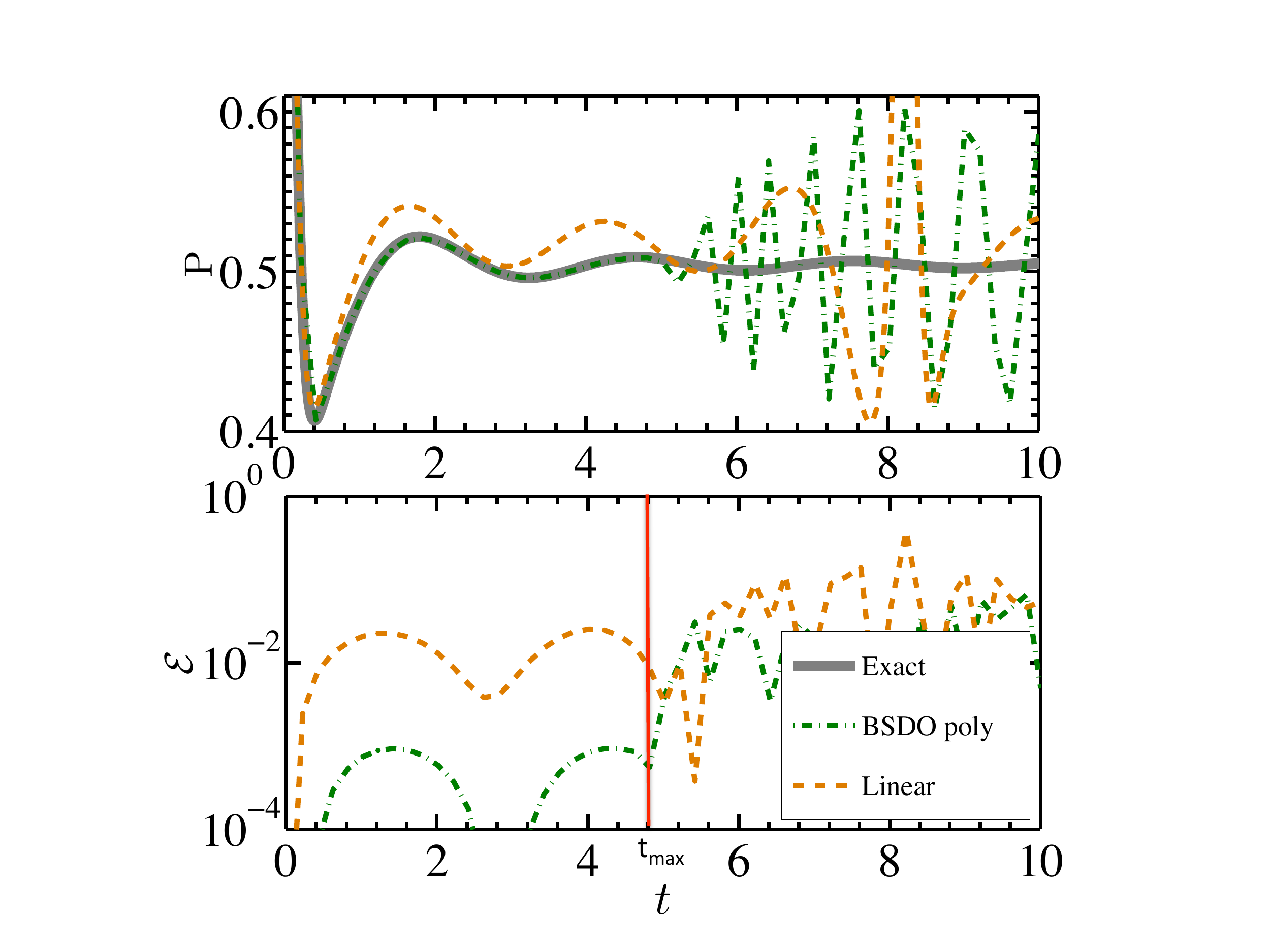}}
\caption{Time evolution of the population of the upper level for $N_b=65$ (upper panel) and error ${\mathcal E}$ according to (\ref{error}) in logarithmic scale (lower panel). In both cases, different discretization schemes are considered. Dot-dashed green and dashed orange curves correspond respectively to polynomial and linear methods. The linear black curve in the upper panel corresponds to the exact solution. The curves and error of the mean and the equal weight method of Sect. (\ref{alternative}) are not shown, but have a similar behaviour as the ones of the linear method. The red line below shows the time $t_{\mmax}$ at which the error of the polynomial method increases two orders of magnitude,which coincides with the exact formula \eqref{tmaxSys} (see also Fig. (\ref{figexp4})). We have considered $\omega_s=0.5$, $\alpha=1$, $s=0.5$, $\omega_c=10$, and a maximum frequency in the spectrum $\omega_\mmax=50$. \label{figexp1}}
\end{figure}

\begin{figure}[t]
\centerline{\includegraphics[width=0.4\textwidth]{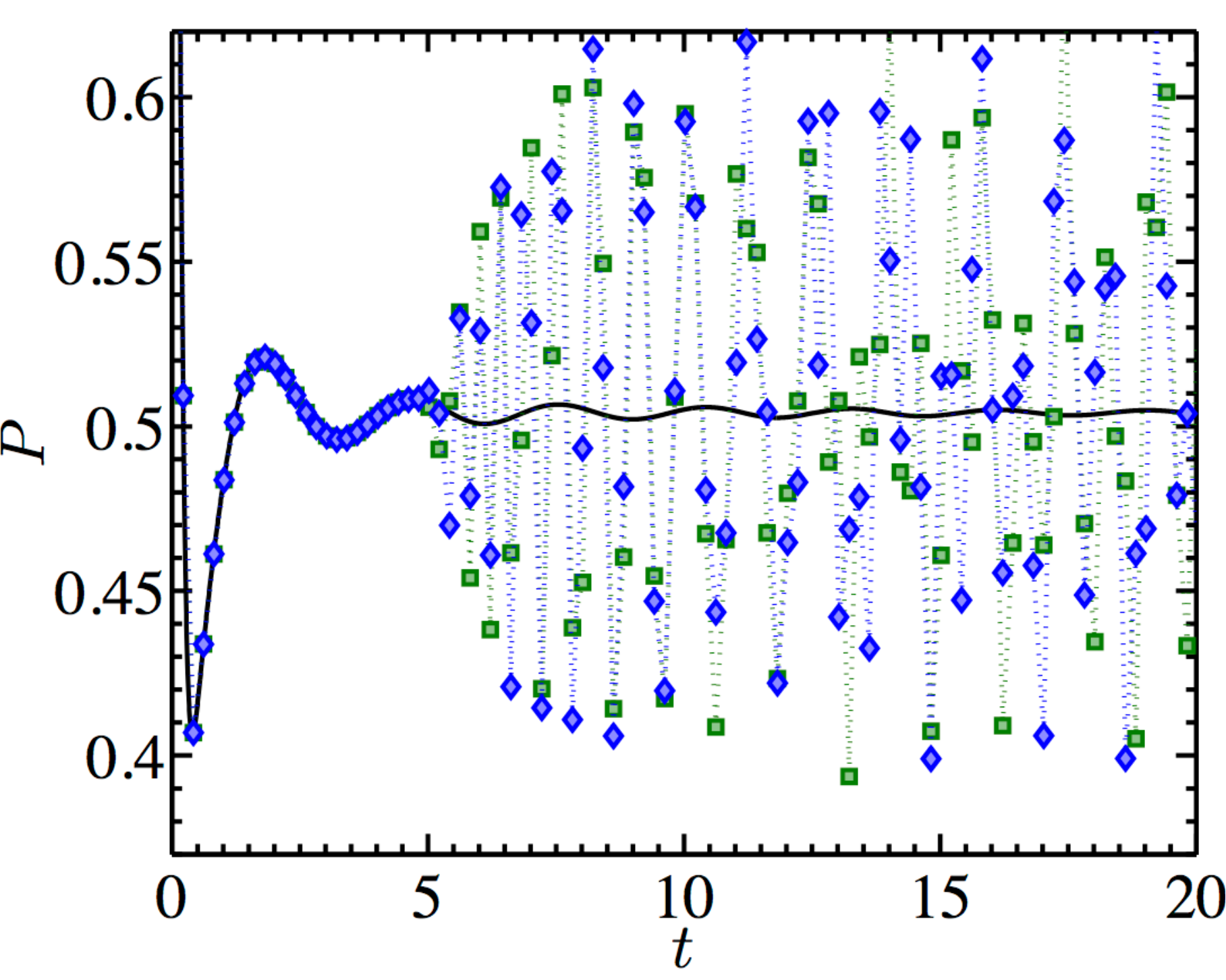}}
\caption{Evolution of the upper level considering the quadrature method with different polynomial classes for $N_b=65$ nodes. 
Blue diamonds, and green squares correspond, respectively, to the Gaussian quadrature rule (with Legendre polynomials), and to the Gauss-Christoffel quadrature with BSDO polynomials (i.e. polynomials obeying the relation (\ref{ortho}) with $w(x)=J(x)$). \label{figexp3}}
\end{figure}

\begin{figure}[t]
\centerline{\includegraphics[width=0.4\textwidth]{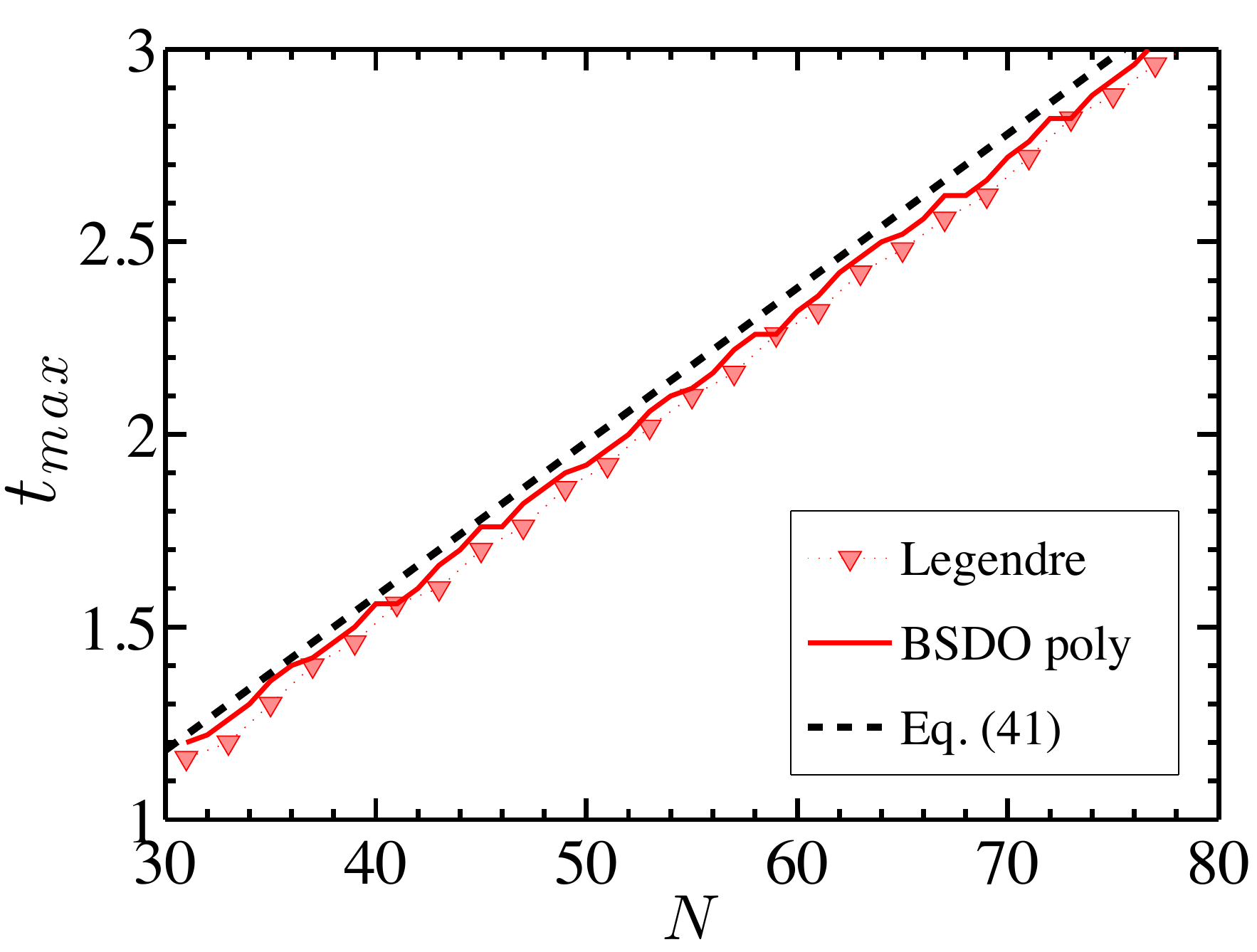}}
\caption{Maximum time at which the error between the evolution with discretization with $N$ nodes, and the exact (continuous) one is below a certain threshold chosen as $0.004$. The maximum frequency in the spectrum is  $\omega_\mmax=100$. Blue diamonds, and green circles correspond, respectively, to the Gaussian quadrature rule (with Legendre polynomials), and to the Gauss-Christoffel quadrature (with BSDO polynomials). 
System parameters are the same as in Fig. (\ref{figexp1}), except for the fact that we are now considering $s=1.5$.~The figure shows approximately the same slope as the one predicted by  eq.~\eqref{tmaxSys}. \label{figexp4}}.
\end{figure}

Considering zero temperature, the OQS dynamics can be easily solved by exact diagonalization (ED), since there is only one excitation involved in the problem (it is a single-particle problem with a quadratic Hamiltonian). In this context, Fig.~(\ref{figexp1}) shows results for the population 
\eq{
P(t) 
& = \langle\sigma^+(t)\sigma^-(t)\rangle \\
{\mathcal E}(t)
& =|P(t)-P\th{discr}(t)|, 
\label{error}
}
where $P(t)$ is computed with the continuous environment, and $P\th{discr}(t)$ is the population computed with the discretized environment.
${\mathcal E}(t)$ is the error made by using the discretized environment. We compare results obtained using 
the linear discretization as an example for a \textit{direct discretization strategy} 
with the \textit{orthogonal polynomial} strategy that uses \eqref{recurrence} 
with the weight function $w(x)=J(x)$ generating BSDO polynomials.
Clearly, the BSDO strategy leads to an error that is at least two orders of 
magnitude smaller than the one of the linear discretization with the same number of modes
until reaching a time $t\tl{tmax}$, when the discretized system fails to accurately describe the continuous system. 
Physically, such a failure can be interpreted as a revival of the system dynamics, 
which occurs when the emitted excitation hits the chain extreme and bounces back into the system.
We note that the results obtained with the heuristic approaches described in Sec.~(\ref{alternative}) (not shown), 
are found to achieve a similar level of accuracy as the linear discretization strategy. 

Fig. (\ref{figexp3})  compares two orthogonal-polynomial based strategies: one generated with \eqref{recurrence} 
using the weight function $w(x)=J(x)$ (BSDO quadrature) and one using  $w(x)=1$ (Legendre quadrature).
The figure confirms the statement made after eq.~\eqref{eqLeg} that both strategies yield basically the same
accuracy if the bath spectral density does not show a severe non-regular behavior. 
Also, as shown in Fig. (\ref{figexp4}), $t_\mmax$ is linearly related to the number of node points considered in the quadrature rule. This follows from equation \eqref{tmaxSys}.

We note that also in the finite temperature case, studied within the second order weak coupling master equation, allows us to recover the result that the BSDO strategy is optimal up to the time $t_{\mmax}$ (see Appendix (\ref{thermal})).  

\begin{figure}[t]
\centerline{\includegraphics[width=0.4\textwidth]{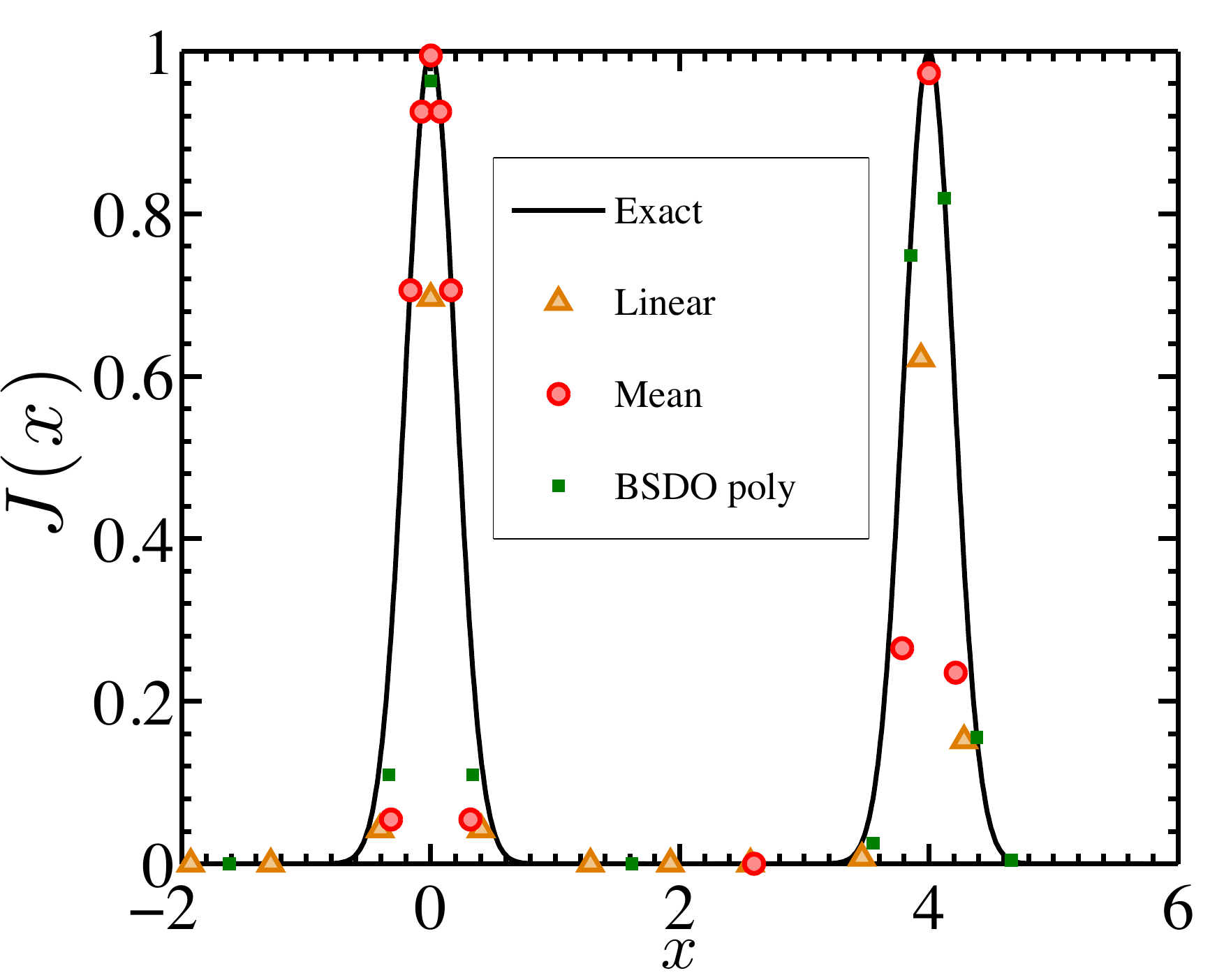}}
\caption{The generic bath spectral density \eqref{Jgauss} and its discretized versions.
To plot the discrete spectral function $J\th{discr}(x)$, we replace the delta
function by a rescaled indicator function $\delta(x-x_n) \rightarrow \chi(x-x_n)/\Delta x_n $,
where $\Delta x_n$ is the width of $I_n = [ (x_n+x_{n-1})/2, (x_{n+1} +x_n)/2]$.
This rescaling accounts for the fact that for comparisons with
the continuous spectral density, the discrete spectral function should be interpreted
as a \textit{probability density} defined on the energy interval $(a,b)$ that associates
a weight (an excitation probability) to an \textit{energy interval}, and not
as a \textit{probability mass function} that associates a weight to a \textit{value} of $x_n$.
}
\label{figJgauss}
\end{figure}

\subsection{Single-impurity Anderson model}

The single-impurity Anderson model (SIAM) has the form of Hamiltonian (\ref{Hcont}), 
with the impurity and bath operators being spin-dependent fermionic 
creation and annihilation operators,
\eq{ \label{HSIAM}
H_\tx{sys} & = U (d_{\uparrow}^\dagger d_{\uparrow} - \frac{1}{2}) (d_{\downarrow}^\dagger d_{\downarrow}-\frac{1}{2}),\\
H\tl{bath}  & = \sum_\sigma \int_{a}^{b} dx\,x\,a_{x\sigma}\dag a_{x\sigma}, \nonumber\\
H\tl{coupl} & = \sum_\sigma \int_{a}^{b} dx\,V(x) (d_\sigma\dag a_{x\sigma} + \text{h.c.}.) \nonumber
}
In a grand-canonical picture this corresponds to the half-filled case
obtained for chemical potential $\mu=-U/2$. The physics of this case 
shows generic features. Clearly, for $U\neq0$, $H_\tx{sys}$ describes a non-quadratic interaction.

The generic case of interest for the physics of strongly-correlated electron systems
is best captured by a bath spectral density of the form
\bea \label{Jgauss}
J(x) & = \sum_{x_0\in\{-4,0,4\}} e^{-\frac{(x-x_0)^2}{2\eta^2}} \tx{ for } x \in [-5,5]
\eea
outside of the interval $[-5,5]$ we set $J(x)=0$. 
This bath spectral density is a superposition of three Gaussian peaks that produces ``gapped"
regions where $J(x)$ is practically zero. Figure \ref{figJgauss}
shows the continuous and the discretized version of this $J(x)$.
The peak at zero frequency corresponds to low-energy excitations in the bath, 
as they are present in a metal. The two other 
peaks correspond to high-energy excitations that become relevant when the interaction
$U$ generates low (single occupation) and high (double or zero occupation) energy states.
In a Mott insulator, there is no low energy physics any more and the interaction created a gap
in the excitation spectrum. The most exciting physics happens in the intermediate
regime where the quantum Mott-Insulator phase transition occurs.

\begin{figure}[t]
\centerline{\includegraphics[width=0.4\textwidth]{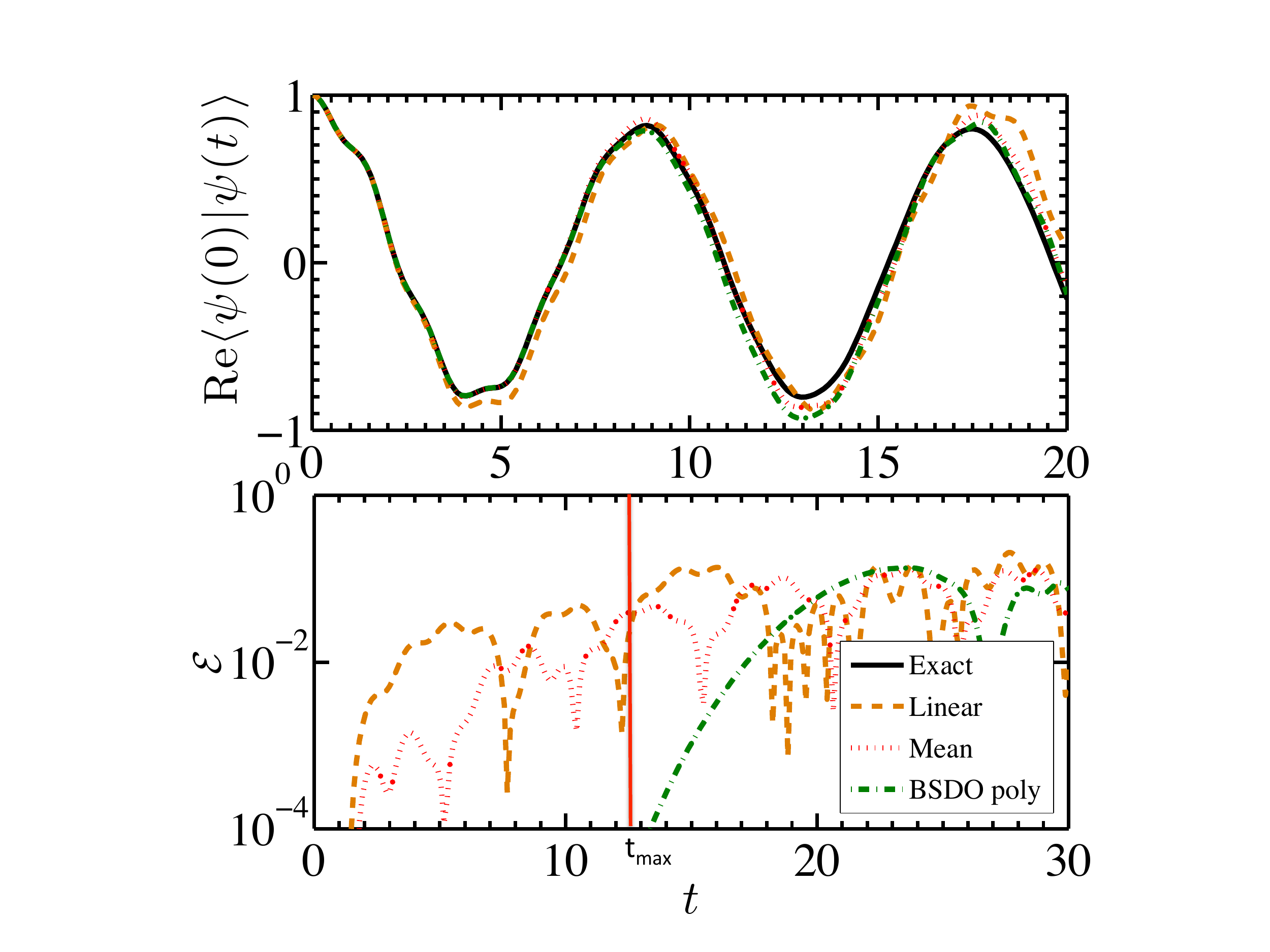}}
\caption{Time evolution of the SIAM \eqref{HSIAM} for $U=0$.
Upper panel: Time evolution for $N_b=15$. 
Lower panel: Error for $N_b=31$.  
The maximal time (red vertical line) until which the BSDO polynomial discretization yields the
exact description can be computed using \eqref{tmaxSys}, 
and yields for $a=-5$, $b=5$ and $N_b=31$ the value $t\tl{max}=12.6$.
}
\label{figSIAMU0}
\end{figure}

\begin{figure}[t]
\centerline{\includegraphics[width=0.4\textwidth]{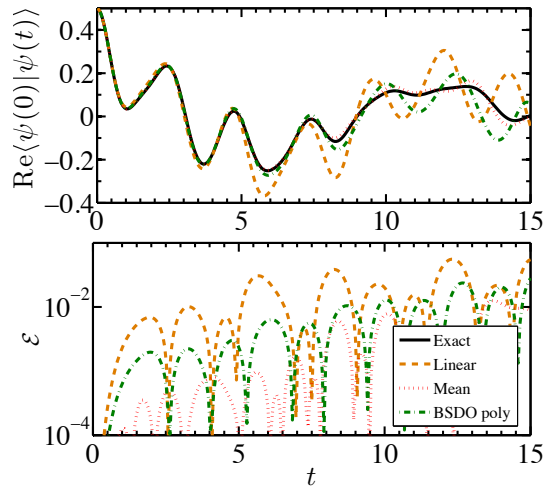}}
\caption{Time evolution of the SIAM \eqref{HSIAM} for $U=4$.
Upper panel: Time evolution for $N_b=15$. 
Lower panel: Error for $N_b=31$. 
}
\label{figSIAMU4}
\end{figure}

Let us first study the non-interacting case $U=0$, which only involves
a quadratic Hamiltonian. In this case, we confirm the 
results of the previous section. Figure \ref{figSIAMU0} shows the 
time evolution of the overlap of the initial state (the Green's function $iG(t)=\bra{\psi_0} e^{-i(H-E_0)t} \ket{\psi_0}$ defined in
\eqref{greenEvol}), that consists in placing a spin-up electron on 
the impurity $\vert \psi(t=0) \rangle = d_{\uparrow}^\dagger\vert E_0 \rangle$,
with its time evolution. Evidently, the linear discretization yields the worst results, and the 
Gauss-Christoffel (BSDO) strategy yields a numerically exact result up to 
time 6.

Let us now turn to the interacting case where $U$ is non-zero and the Hamiltonian is no longer quadratic. 
Figure \ref{figSIAMU4} confirms the result of Sec.~\ref{timeEvol}
that BSDO polynomials do no longer give optimal results as they no longer generate a Gaussian quadrature rule.
Now the heuristic \tit{mean} method produces the best results, leading to errors that are at least a factor 2 
smaller than the BSDO strategy. The mean method directly uses the fact that one can ignore
gapped regions in the bath spectral density. This is important in the computation of strongly correlated materials.
In both cases described in Figs. \ref{figSIAMU0} and \ref{figSIAMU4}, the equal weight method of Sec. \ref{alternative} performs qualitatively similar to the mean method, and therefore it has not been shown for the shake of clarity in the figure.

\section{Conclusions}
\label{conclusion}

In this paper we have analyzed a OQS coupled to a bosonic environment characterized by a Caldeira and Leggett type of spectral density, and a quantum impurity model consisting on an impurity coupled to a fermionic bath. 
We considered \textit{direct discretization strategies} and \textit{orthogonal polynomial} quadrature based strategies. 
We have shown that when using orthogonal polynomials, the choice of the polynomial class does 
not affect considerably the error in the resulting system dynamics.
In addition, we have shown that the Gauss-Christoffel quadrature rule (which is based on the choice of a particular family of polynomials here denoted as BSDO), correspond to the chain mapping approach proposed by Refs.~\onlinecite{prior2010,chin2011}. Such chain mapping leads effectively to a discrete chain representation, 
which when transformed back to a diagonal form, leads to environment eigenvalues 
that precisely correspond to the nodes of the Gauss-Christoffel (BSDO) quadrature rule.

Finally, we have shown that in a non-interacting system (i.e.~with quadratic Hamiltonian), the polynomial quadrature method is exact at short times. Nevertheless, for non-quadratic Hamiltonians (like an an impurity with non-zero interaction term) this is no longer the case. This means that the notion of optimality that is associated with an optimal representation of the continuous integral of $J(x)$ by a finite number of points breaks down if we consider non-quadratic Hamiltonians. In other words, the non-linear problem that is encoded in such non-quadratic Hamiltonian obviously will no longer be well described by just considering a polynomial quadrature rule on the integral. It is noted that, although we have presented a scheme (the mean method) that performs better than Gauss Christoffel quadrature in this case, we showed that a general statement \textit{cannot} be made.

Note added in proof. Dynamical error bounds on expectation values of system observables for a Hamiltonian discretised using orthogonal polynomials have recently been derived in Ref. \cite{woods2015}.

\section{Acknowledgements}

We thank G. K.-L. Chan and N.-O.~Linden for useful discussions.
We acknowledge the use of I.~P.~McCulloch's 
DMRG code for the calculation of the interacting SIAM, the
results of which are shown in Fig.~\ref{figSIAMU4}.
IDV would like to thank Nanosystems Initiative Munich (NIM) (project No. 862050-2) for support, as well as the Spanish MICINN (Grant No. FIS2013-41352-P) for partial support.
FAW and US acknowledge funding from \href{https://for1807.physik.uni-wuerzburg.de/}{FOR1807} of the DFG.

\appendix

\section{Numerical optimization}
\label{secOpt}

Numerical optimization can be formulated in a straightforward way 
for the hybrdization function $\Lambda(z)$, defined in \eqref{Lambda},
evaluated on a grid of imaginary frequencies $z=i\omega_k$,
\eq{ \label{imagCost}
\chi^2 = \sum_k |\Lambda(i\omega_k) - \Lambda\th{discr}(i\omega_k)|^2,
}
using standard numerical minimization techniques  \cite{caffarel1994,go15,liebsch12}.
On the real axis, the equivalent cost function can be formally defined as
$\chi^2 =\sum_k |\Lambda(\omega_k+i0^+)- \Lambda\th{discr}(\omega_k+i0^+) |^2$, 
but is of no use as the difference of a continuous function
and a singular function is always infinite. Therefore, we \textit{cannot} use
numerical optimization to discretize the 
spectral representation of the continuous bath,
i.e.~the hybridization function evaluated on the real 
axis via $J(x)= -\frac{1}{\pi} \tx{Im} \Lambda(\omega+i0^+)$.

If one carries out the optimization on the imaginary axis via \eqref{imagCost}, 
one obtains a set of parameters $\{x_n,V_n\}$ for the discrete bath and an 
associated hybridization function $\Lambda^\text{discr}(z)$,
which gives a quantitatively precise approximation to $\Lambda(z)$ \tit{only} 
when evaluated on the imaginary frequency axis. On the real-frequency axis, 
the approximation is very rough and can only be considered qualitatively correct.
This follows already from the fact that only relatively small numbers 
of bath sites $N_b\lesssim 15$ can be stably optimized.
Still the approach is valid if one is satisfied with the much lower precision 
on the real axis and does not strive to describe real-time evolution as in this paper.
The preceding statements are \eg discussed, among several other results, 
in \oc{wolf15i}, where the goal was not to describe real-time evolution but 
``thermodynamic'' properties.

We note that one \textit{can} define a meaningful cost function on the real axis, 
if one allows for non-hermitian Hamiltonians with complex bath energies,
or an equivalent description in terms of Lindbladt operators \cite{dorda2014}.

We further note that one can also construct an \tit{optimal} discrete representation
of the ``second bath" that appears within non-equilibrium DMFT \cite{gramsch13}. But this only suffices
to describe situations in which the system and bath are initially 
not entangled \cite{gramsch13}. As non-equilibrium DMFT is a promising
approach to describe the non-equilibrium dynamics of strongly correlated
materials, it is desirable to extend the promising DMRG calculations
for situations with a non-entangled initial state \cite{wolf14i,balzer15} to 
the general case of entangled initial states. But then one also has to discretize
the ``first" bath, which incorporates the spectral information of $H$ and which is 
equivalent to the bath that is the subject of the present paper. For the
first bath, one again faces the problem that a cost function cannot be
meaningfully defined.

\section{System Green's function}
\label{secGreen}

The retarded system Green's function 
is defined in terms of the general 
retarded Green's function (system and bath)
\eq{
G(x) = \frac{1}{x + i0 - (H-E_0)}
}
by taking expectation values
with respect to the system states \cite{hewson2003},  
\eg $\ket{\psi_0} = d\dag \ket{E_0}$
\eq{
G\tl{sys}(x) = \bra{\psi_0} G(x) \ket{\psi_0}.
}
For the system Hamiltonian $H\tl{sys} = \epsilon_0 d^\dagger d$ it reads \cite{hewson2003}
it can be evaluated as
\eq{
G\tl{sys}(x) = \frac{1}{x + i0 - \epsilon_0 + \Lambda(x)}.
}
where $\Lambda(x)$ is defined in \eqref{Lambda}.

\section{Lanczos algorithm}
\label{secLanczos}
\subsection{General Lanczos algorithm and relation to orthogonal polynomials}
\label{secLanczosGen}

The Lanczos algorithm constructs a three-diagonal matrix representation  
of any Hermitian operator $H$ by representing it 
in its Gram-Schmidt orthogonalized Krylov basis $\{\ket{f_n}\}$:
Given a start vector $\ket{f_0}$ that has non-zero overlap with all eigen-states
of $H$, one orthogonalizes the vector $\ket{f_n}$  
with respect to all previous vectors $\ket{f_{n'}}$ with $n'<n$. This results in
\eq{\label{lanczosGen}
\alpha_n
& = \bra{f_n} H \ket{f_n}, \nonumber\\
\ket{r_{n}}
& = H \ket{f_n} - \alpha_n \ket{f_n} - \sqrt{\beta_{n}} \ket{f_{n-1}}  \nonumber\\
\beta_{n+1}
& = \abs{\ip{r_{n}}{r_{n}}}, \quad \beta_{0} = 0  \nonumber\\
\ket{f_{n+1}}
& = \frac{1}{\sqrt{\beta_{n+1}}} \ket{r_n}, \quad \tx{for}~n = 0, \dots,  N-1.
}

One can show that the Lanczos algorithm implicitly constructs
a family of polynomials $q_n(x)$ that are orthogonal with respect to 
an inner product weighted with the spectral density $A(x)$ of the operator $H$ \cite{gautschi2005,justiniano15}
\eq{
w(x) = \sum_{n=1}^{\tx{dim}(H)} |\ip{E_n}{f_0}|^2 \, \delta(E-E_n) = A(x). \nonumber
}
The proof is as follows. Let us define the polynomial $q_n(x)$ of degree $n$ via 
\eq{
\ket{f_{n}} = q_{n}(H) \ket{f_0}, \label{qnlanczos}
}
and then show that they are orthogonal with respect to $A(x)$. 
We note that \eqref{qnlanczos} can always be fulfilled as $\ket{f_{n}}$ is constructed by 
applying $H$ $n$ times to the initial state $\ket{f_0}$. Furthermore,
\eq{
\int_a^b dx & A(x) q_k(x) q_l(x) = \sum_{n=1}^{N_b}  \ip{f_0}{E_n} q_k(E_n) q_l(E_n) \ip{E_n}{f_0} \nonumber\\
& = \bra{f_0} q_k(H) q_l(H) \ket{f_0} = \ip{f_{k}}{f_{l}} = \delta_{kl}, \nonumber
}
which completes the proof.

\subsection{Chain mapping}

In the following, we show how to use the Lanczos algorithm to tridiagonlize the star Hamiltonians 
$H$ in \eqref{Hcont} and $H\th{discr}$ in \eqref{Hdiscr}. This amounts
to using the general algorithm \eqref{lanczosGen} for the bath Hamiltonians
$H\tl{bath}$ and $H\tl{bath}\th{discr}$, respectively. The bath Hamiltonians
are quadratic and therefore simple to treat. They have the spectral 
densities $J(x)$ and $J\th{discr}(x)$ as defined in \eqref{Jcont} and \eqref{Jdiscr}, respectively.
Already from this we can conclude from the argument of Sec.~\ref{secLanczosGen}, 
that the Lanczos algorithm applied for the continuous $H\tl{bath}$, 
yields the same set of orthogonal polynomials as the recurrence \eqref{recurrence},
and is therefore equivalent to it.

In practice, the algorithm is usually used to obtain representations of the 
discrete bath and coupling Hamiltonians $H\tl{bath}\th{discr}$
and $H\tl{coupl}\th{discr}$. We will lay out the procedure
for the discrete case, and note differences to the continuous case where
necessary.

Let us denote the (single-particle) bath orbital states of the discrete star 
representation \eqref{Hdiscr} as $\ket{c_n}$. These are associated 
with the operators $c_{n}\dag$ via $\ket{c_n} = c_{n}\dag\ket{\tx{vac}}$. 
Analogously, define the bath orbitals of the 
chain representation \eqref{map} as $\ket{e_n}$ where $\ket{e_n} = e_{n}\dag\ket{\tx{vac}}$. 
The first orbital of the chain representation then is
\eq{ \lb{eqFirstChainState}
\ket{e_{0}} 
&= \frac{1}{V_\tx{tot}} \sum_{n=1}^{N_b} V_{n} \ket{c_{n}} ,  \\  
& |V_\tx{tot}|^2 
= \sum_{n=1}^{N_b}  \abs{V_{n}}^2 = \int_a^b dx\,J(x), \nonumber
}
in the discrete case, and 
\eq{ \lb{eqFirstChainStateCont}
\ket{e_{0}} 
&= \frac{1}{V_\tx{tot}} \int_a^b dx\, V(x) \ket{a_{x}},  \quad \ket{a_{x}} = a_x\dag \ket{\text{vac}},
}
in the continuous case, in agreement with \eqref{enoperators}. 
In both cases, it is a superposition of all states in the star.
The coupling Hamiltonians 
$H\th{discr}\tl{coupl}$ in \eqref{Hdiscr} 
can then be written as $H\th{discr}_\tx{coupl} = V\tl{tot} ( \ket{d}\bra{e_{0}} + \tx{h.c.})$,
where $\ket{d}$ is associated with the system operator $d\dag$.
The same equation holds in the continuous case.

One then uses the Lanczos algorithm to construct a three-diagonal representation 
of $H\th{discr}\tl{bath}$
\eq{\label{lanczos}
\alpha_n
& = \bra{e_n} H\th{discr}\tl{bath} \ket{e_n}, \\
\ket{r_{n}}
& = H\th{discr}\tl{bath} \ket{e_n} - \alpha_n \ket{e_n} - \sqrt{\beta_{n}} \ket{e_{n-1}}  \nonumber\\
\beta_{n+1}
& = \abs{\ip{r_{n}}{r_{n}}}, \quad  \beta_{0} = 0 \nonumber\\
\ket{e_{n+1}}
& = \frac{1}{\sqrt{\beta_{n+1}}} \ket{r_n}, \quad \tx{for}~n = 0, \dots,  N_b-1. \nonumber
}
or analogously, for the continuous case. The parameters $\alpha_n$ and $\beta_n$ in the recursion
are the parameters of the Hamiltonian \eqref{map}, and with that the map is complete.

In practice we note that we \textit{cannot} find a direct matrix representation
of the continuous Hamiltonian \eqref{Hcont} that we could 
use on a computer to compute \eqref{lanczos}. 
In the discrete case, on the other hand,
the preceding equations are easily solved by generating
a matrix representation by multiplying 
from the left with $\bra{c_{n'}}$ and inserting 
identities $\sum_{n'} \ket{c_{n'}}\bra{c_{n'}}$ such that
the initial vector can be written as 
$(\ip{c_n}{e_{0}})_{n=1}^{N^b} = (V_n)_{n=1}^{N_b}$
and the representation of $H\th{discr}\tl{bath}$ involved is 
$\bra{c_n} H\tl{bath} \ket{c_{n'}} = x_n \d_{nn'}$. 

Due to the numerical
instability of the Lanczos algorithm, the recurrences \eqref{lanczos} and \eqref{recurrence} have to be 
computed with high-precision arithmetics when exceeding
$N_b \sim 40$ or using the stabilized implementation of \oc{gautschi2005}.

\begin{figure}[t]
\centerline{\includegraphics[width=0.4\textwidth]{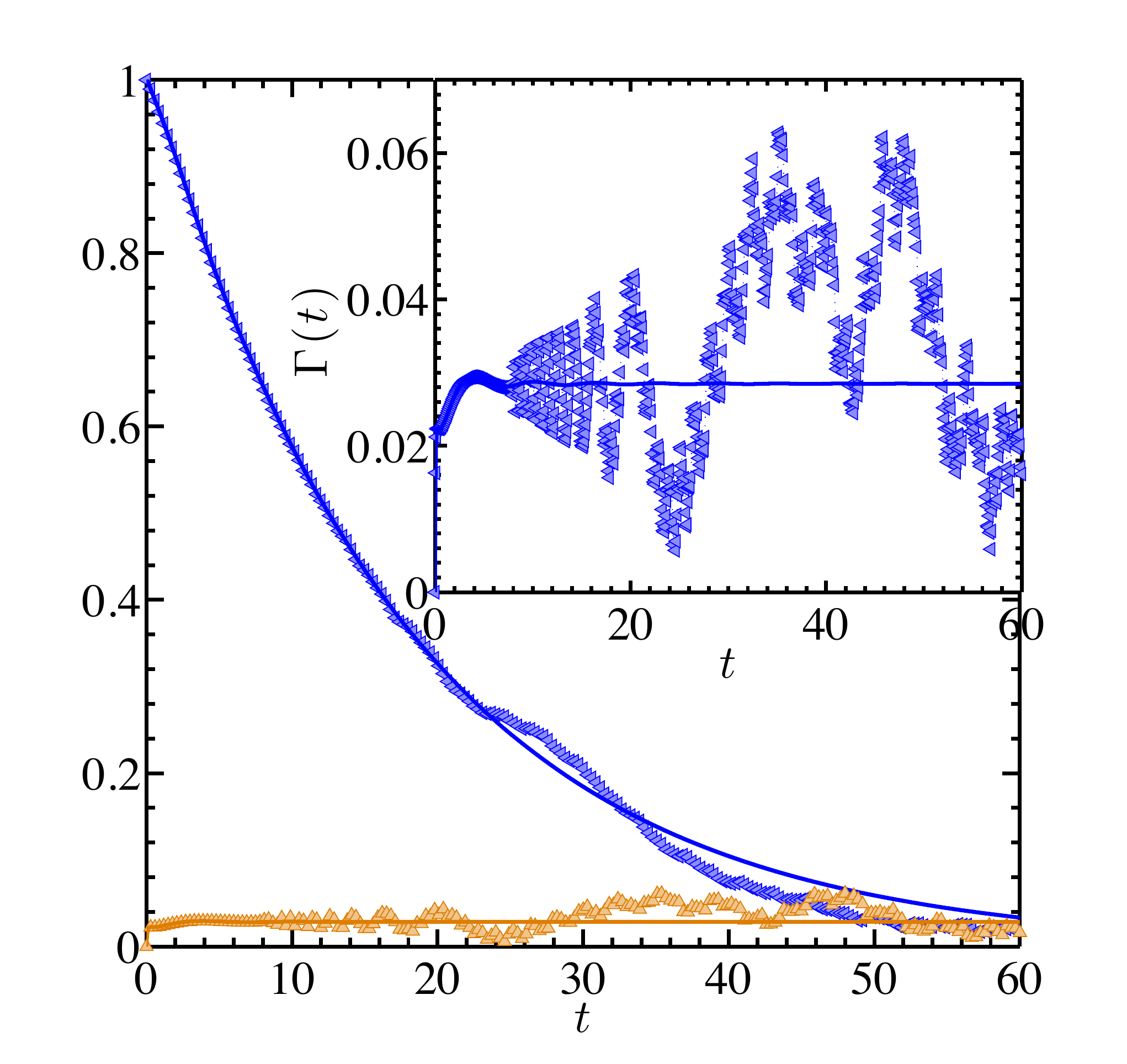}}
\caption{Blue curves represent the master equation solution for the atomic population $\langle \sigma^\dagger (t)\sigma(t)\rangle$, with quasi-continuous spectrum (plain solid curve) and Gauss-Christoffel (BSDO) quadrature with $N=100$ nodes (curve with triangles). Orange curves (see also inset) represent the evolution of $\Gamma(t)=\int_0^t \alpha_T(\tau)e^{i\omega_S \tau}$ for the same two cases. The spectral density, as well as all system parameters are the same as in Fig. (\ref{figexp1}), except for the coupling that now is considered to be weak, $\alpha=0.01$.\label{figexp6}}
\end{figure}

\section{Estimate the error in time evolution}
\label{secChebychev}

As Chebyshev polynomials are \textit{almost optimal} they will result in a sequence $c_n$,
which is very close to the sequence produced by an optimal choice of polynomials, 
in the sense of the discussion of \eqref{expand}.

For Chebyshev polynomials ($v(x) = \wt v(x') = \frac{1}{\pi} (1-x')^{-\frac{1}{2}}$
and $q_n(x)=\wt q_n(x') = \arccos(n \cos(x'))$ with $x'=2\frac{x-a}{b-a}-1$, $x=\frac{1}{2}(b-a) x' + \frac{1}{2} (b+a)$), 
we can evaluate the coefficients in \eqref{expand} explicitely,
\eq{
c_n 
& = \tfrac{2}{b-a} e^{-\frac{i}{2} (b+a) t} \int_{-1}^{1} dx' \, \wt v(x') e^{-\frac{i}{2}(b-a)t} \wt q_n(x) \nonumber \\
& =  \tfrac{2(-i)^n}{b-a} e^{-\frac{i}{2} (b+a) t}  J_n\big(\tfrac{1}{2}(b-a) t\big),
}
where $J_n(t')$ are Bessel functions of the first kind. For all practical purposes,
$J_n(t') \simeq 0$ if $n>t'$. More concretely, the asymptotic form
for high values of $n$ reads $n \gg t'^2-1$,  $J_n(t') \sim \frac{1}{(n+1)!} (\tfrac{t'}{2})^n$ \cite{abramowitz65},
and shows that this decreases as a faculty.

\section{Open quantum system in the presence of a thermal environment}
\label{thermal}

\begin{figure}[t]
\centerline{\includegraphics[width=0.4\textwidth]{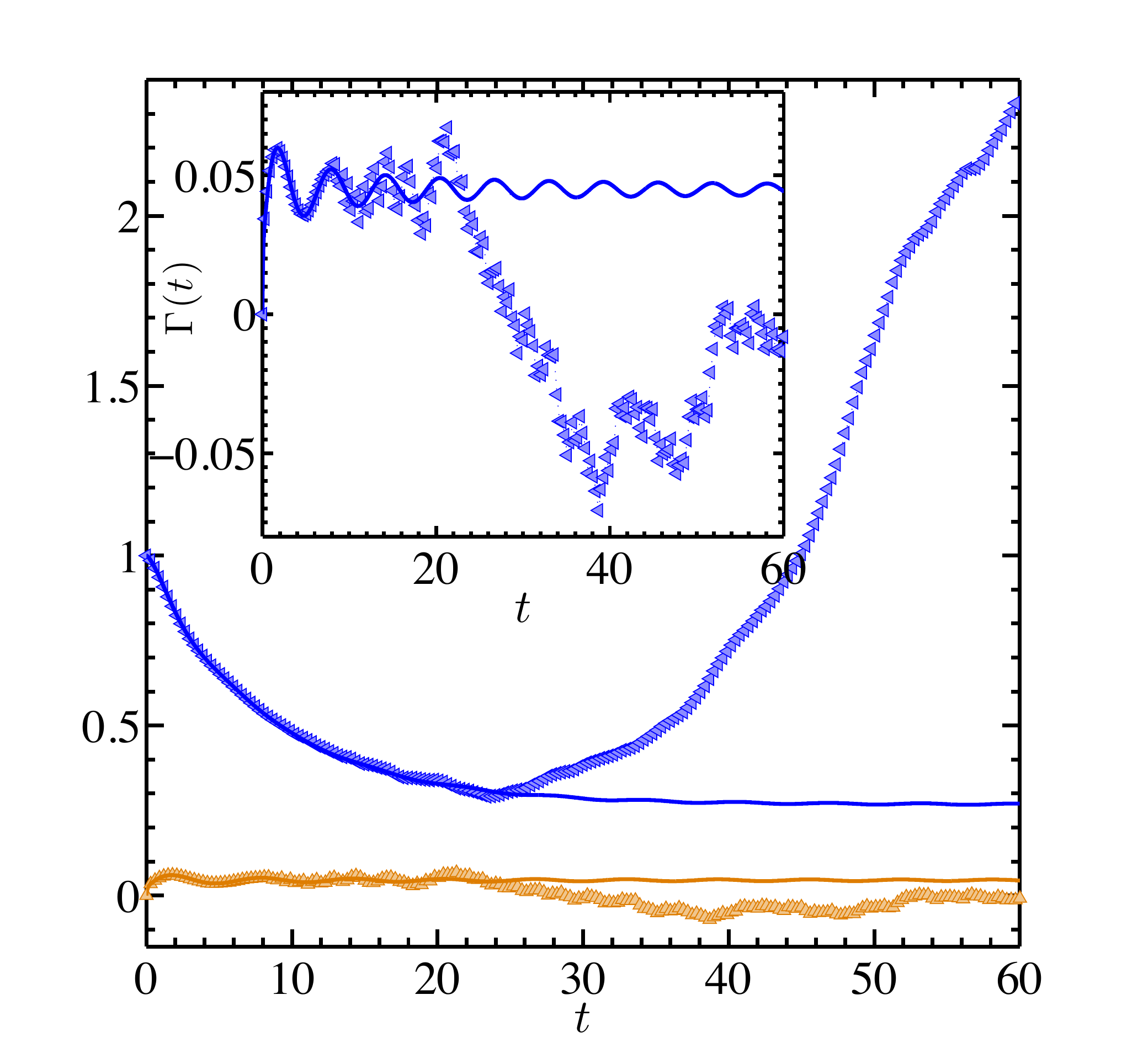}}
\caption{Same as in Fig. (\ref{figexp6}), but considering finite temperature ($\beta=1$).\label{figexp7}}
\end{figure}

Let us now consider $d=d^\dagger=\sigma_x$ in (\ref{Hcont}), and a finite temperature in the environment. We study this case within a standard approximate scheme, namely a master equation (ME) up to second order in the system-environment coupling parameter $g$ 
\cite{breuerbook},
\begin{eqnarray}
\frac{d\rho_s  (t)}{dt}&=&-i[H\tl{sys} ,\rho_s (t) ]+\int_0^t d\tau \alpha_2^{*}(t-\tau) [d^\dagger ,\rho_s (t)  d(\tau-t)]\nonumber\\
&+&\int_0^t d\tau \alpha_2(t-\tau) [d^\dagger(\tau-t) \rho_s (t) ,d]\nonumber\\
&+&\int_0^t d\tau\alpha_1(t-\tau)[d(\tau-t)\rho_s (t) ,d^{\dagger}]\nonumber\\
&+&\int_0^t d\tau \alpha_1^{*} (t-\tau)[d,\rho_s (t) d(\tau-t)^{\dagger}]+{\mathcal O}(g^3),
\label{icc20chap3a}
\end{eqnarray}
with $\alpha_1 (t-\tau)=\sum_k g^2_k (n_k+1)e^{-i\omega_k (t-\tau)}$, $\alpha_2 (t-\tau)=\sum_\lambda g^2_k n_ke^{i\omega_k (t-\tau)}$, and $d(t)=e^{iH\tl{sys} t}de^{-iH\tl{sys} t}$. 

As it can be seen in Figs.~\ref{figexp6}, for zero temperature, and \ref{figexp7} for finite temperature, the polynomial Gauss-Christoffel (BSDO) quadrature is still extremely accurate a short times. Nevertheless, just as in the zero temperature case, after a certain time $t_\mmax$, the discretization procedure starts to fail. Such a failure is originated from the fact that the polynomial quadrature rule starts to reproduce inaccurately the integrals $\Gamma(t)=\int_0^t \alpha_T(\tau)e^{i\omega_S \tau}$, with $\alpha_T(t)=\alpha_1(t)+\alpha_2^*(t)$, entering in the master equation. 

Indeed, as seen in the inset of both figures, small deviations of this quantity due to an inaccurate discretization, produce large deviations in the dynamics with respect to the reference (corresponding to the solution with a quasi-continuous spectrum), and this deviation is particularly large at finite temperatures.

\bibliography{Bibtexelesdrop_17}

\end{document}